\documentclass[preprint,10pt,times]{elsarticle}
\usepackage[a4paper, margin=1.3in]{geometry} %
\usepackage{setspace}
\usepackage{amsmath,amsfonts}
\usepackage{amssymb}
\usepackage{mathrsfs}
\usepackage{lipsum}
\usepackage{graphicx}%
\usepackage{textcomp}
\usepackage{multirow}
\usepackage{multicol}
\usepackage{algpseudocode}
\usepackage[switch]{lineno}
\usepackage{xcolor}
\usepackage{flushend}
\usepackage{hyperref}
\hypersetup{
    colorlinks=true, 
    linkcolor=blue,  
    citecolor=blue   
}

\begin{document}
\setstretch{1.2}
\begin{frontmatter}

\title{CT-Mamba: A Hybrid Convolutional State Space Model for Low-Dose CT Denoising}

\author[label1]{Linxuan Li}
\author[label1]{Wenjia Wei}
\author[label1]{Luyao Yang}
\author[label1]{Wenwen Zhang}
\author[label1,label4]{Jiashu Dong}
\author[label2]{Yahua Liu}
\author[label3]{Hongshi Huang}
\author[label1,label4,label5]{Wei Zhao\corref{cor1}}
\cortext[cor1]{Corresponding author.}

\affiliation[label1]{organization={School of Physics, Beihang University},
            city={Beijing},
            country={China}}

\affiliation[label2]{organization={Emergency Department of the Third Medical Center, Chinese PLA General Hospital},
            city={Beijing},
            country={China}}

\affiliation[label3]{organization={Department of Sports Medicine, Peking University Third Hospital, Institute of Sports Medicine of Peking University, Beijing Key Laboratory of Sports Injuries, Engineering Research Center of Sports Trauma Treatment Technology and Devices, Ministry of Education},
            city={Beijing},
            country={China}}

\affiliation[label4]{organization={Hangzhou International Innovation Institute, Beihang University},
            city={Hangzhou},
            country={China}}

\affiliation[label5]{organization={Tianmushan Laboratory},
            city={Hangzhou},
            country={China}}

\begin{abstract}
Low-dose CT (LDCT) significantly reduces the radiation dose received by patients, however, dose reduction introduces additional noise and artifacts. Currently, denoising methods based on convolutional neural networks (CNNs) face limitations in long-range modeling capabilities, while Transformer-based denoising methods, although capable of powerful long-range modeling, suffer from high computational complexity. Furthermore, the denoised images predicted by deep learning-based techniques inevitably exhibit differences in noise distribution compared to normal-dose CT (NDCT) images, which can also impact the final image quality and diagnostic outcomes. This paper proposes CT-Mamba, a hybrid convolutional State Space Model for LDCT image denoising. The model combines the local feature extraction advantages of CNNs with Mamba's strength in capturing long-range dependencies, enabling it to capture both local details and global context. Additionally, we introduce an innovative spatially coherent Z-shaped scanning scheme to ensure spatial continuity between adjacent pixels in the image. We design a Mamba-driven deep noise power spectrum (NPS) loss function to guide model training, ensuring that the noise texture of the denoised LDCT images closely resembles that of NDCT images, thereby enhancing overall image quality and diagnostic value. Experimental results have demonstrated that CT-Mamba performs excellently in reducing noise in LDCT images, enhancing detail preservation, and optimizing noise texture distribution, and exhibits higher statistical similarity with the radiomics features of NDCT images. The proposed CT-Mamba demonstrates outstanding performance in LDCT denoising and holds promise as a representative approach for applying the Mamba framework to LDCT denoising tasks. 
\end{abstract}


\end{frontmatter}


\section{Introduction}
\label{introduction}

\footnotetext[1]{Linxuan Li and Wenjia Wei contributed equally to this work.}
\footnotetext[2]{\href{https://github.com/linxuan-li/CT-Mamba}{https://github.com/linxuan-li/CT-Mamba}}
\footnotetext[3]{Email: \href{mailto:linxuan.li@163.com}{linxuan.li@163.com}, \href{mailto:weiwenjia@buaa.edu.cn}{weiwenjia@buaa.edu.cn}}
\footnotetext[4]{DOI: \href{https://doi.org/10.1016/j.compmedimag.2025.102595}{10.1016/j.compmedimag.2025.102595}}

Computed tomography (CT) is an essential imaging technique in clinical practice, providing crucial anatomical information that aids physicians in making appropriate medical decisions~\citep{zhang2025data, lv2024qualitative}. However, frequent CT scans can significantly increase the radiation dose received by patients, particularly in radiation therapy, where multiple scans are often required to achieve precise tumor localization and minimize damage to organs at risk. Excessive radiation may harm the patient, impact organ function, increase the incidence of radiation-related diseases linked to genetic damage, and ultimately reduce the patient’s quality of life. Therefore, reducing the radiation dose of CT imaging has become a focus of attention.  Low-dose CT (LDCT) is an effective method for reducing patient radiation exposure~\citep{bosch2023risk}. In clinical practice, to acquire LDCT images, it is common to reduce the tube current to decrease the flux of the electron beam that generates X-ray photons, and/or reduce the number of projections during CT procedures. While these methods reduce radiation dose, they also introduce additional noise and artifacts, which decrease the overall quality of the image. If these issues are not effectively addressed, they will significantly impact the application of LDCT in various clinical scenarios. How to obtain high-quality CT image`s that are comparable to normal-dose CT (NDCT) images and meet the clinical utilities from low-dose scanning protocols has become a long-standing and practically significant problem in the field of CT imaging.

To address this issue, several LDCT imaging algorithms have been proposed, which can be roughly divided into the following three categories~\citep{zhang2024review, li2025ddoct, zhang2021noise2context}: sinogram-based preprocessing, iterative reconstruction, and image post-processing. 
Sinogram-based preprocessing methods directly target the raw data acquisition stage and designs specific filters for processing CT projection data obtained through low-dose X-rays. Typical methods include bilateral filtering~\citep{manduca2009projection}, structure-adaptive filtering~\citep{balda2012ray}, and penalized weighted least squares~\citep{wang2006penalized}. These denoising methods  combine physical characteristics with photon statistical characteristics, making them relatively simple and easy to implement. However, these methods rely heavily on high-quality original projection data, limiting their ability to effectively restore undersampled or missing signals, and their practical application is limited by the difficulty in obtaining complete sinogram data. Iterative reconstruction methods~\citep{chen2014artifact} typically reconstruct images by iteratively optimizing an objective function. It usually alternates between forward and backward projections in the projection domain and the image domain until the objective function is minimized based on the convergence criterion. However, this method typically requires access to the raw data and bears a high computational cost. In addition, the quality of image reconstruction heavily relies on the precise setting of the objective function and hyperparameters. These limitations hinder the application of iterative reconstruction methods in clinical practice.

In contrast, image post-processing methods primarily address noise and artifacts within the image domain. Typical methods include non-local means filtering algorithm~\citep{li2014adaptive}, and block matching algorithm~\citep{kang2013image}. Compared to the first two types of methods, post-processing has the advantage of not requiring original projection data from the CT vendor, making it easier to integrate into clinical CT imaging workflows.

In recent years, with the continuous improvement of computer performance, and the rapid development of cloud computing and distributed computing,  deep learning-based image post-processing methods have achieved remarkable achievements in the field of LDCT denoising. Compared with traditional post-processing approaches, deep learning models are capable of automatically learning complex patterns in images, thereby exhibiting superior denoising performance. Currently, two popular network architectures, convolutional neural networks (CNNs)~\citep{lecun1995convolutional} and Transformers~\citep{vaswani2017attention}, are widely applied in LDCT denoising. Furthermore, hybrid models combining these two architectures, along with other neural network frameworks, are continually being developed.

CNN-based methods effectively extract image features through convolutional layers,  and have achieved remarkable success in tasks such as image denoising~\citep{kurmi2024enhancing, kumar2022cnn}, object detection~\citep{redmon2016you}, and image segmentation~\citep{ronneberger2015u}. In recent years, CNNs have also been widely applied in medical image processing tasks, such as automated detection of knee joint synovial fluid in MR images~\citep{iqbal2020deep}, multi-class classification of skin lesions in dermoscopic images~\citep{iqbal2021automated}. In the field of LDCT denoising, several representative CNN-based models have likewise been proposed~\citep{ming2020low, li2022low, gholizadeh2020deep, kyung2024generative, meng2024ddt, zhang2024multi, ko2024adapting}.  
However, due to the limited receptive field of convolutional operators, CNNs exhibit certain limitations in modeling long-range dependencies and capturing global semantic information in images.
Originally designed for natural language processing, Transformer has been successfully adapted to computer vision tasks, where its self-attention mechanism demonstrates remarkable capabilities in modeling long-range dependencies and capturing global semantic information in images~\citep{10990033}. Vision Transformer was the first to apply this architecture to image classification, achieving performance comparable to CNNs and laying the foundation for the broader adoption of Transformers in visual tasks~\citep{dosovitskiy2020image}.
Swin Transformer improved computational efficiency by introducing a hierarchical architecture and shifted window attention, gradually evolving into a general-purpose backbone network for various vision tasks. However, Transformer-based models often incur high computational costs due to the quadratic complexity of the self-attention mechanism with respect to the number of input tokens. This characteristic presents a significant bottleneck in high-resolution image processing and edge computing scenarios.

Diffusion models have achieved remarkable results  in visual tasks in recent years~\citep{yang2023diffusion}. Representative works, for example, the Denoising Diffusion Probabilistic Models (DDPM), proposed by Ho et al., define a forward process that gradually corrupts an image by adding Gaussian noise, and train a noise prediction network (typically a U-Net) to approximate the reverse denoising process, thereby progressively reconstructing a clean image from noise~\citep{ho2020denoising}. On this basis, various extensions have been proposed to improve modeling efficiency. For instance, Denoising Diffusion Implicit Models (DDIM) accelerate the generation process by constructing a deterministic sampling path with non-Markovian properties~\citep{song2021denoising}. Latent Diffusion Models (LDM) perform the diffusion process in a pretrained low-dimensional latent space, significantly reducing the computational burden for high-resolution image generation~\citep{Rombach_2022_CVPR}.
In the field of LDCT denoising, studies have demonstrated the effectiveness of diffusion models in suppressing noise artifacts and improving image quality~\citep{xia2025cmc, ma2025convergent}.
However, current diffusion models generally suffer from slow inference speed and high computational cost, which also limit their deployment in scenarios such as edge computing~\citep{kebaili2025multi, wang2024review}. To overcome these limitations, researchers are working to improve these models.

In this context, the structured state space sequence models (S4) have attracted extensive attention due to their efficient performance in long-range modeling~\citep{gu2022efficiently}. S4 optimizes its internal state representation and information transmission mechanism, significantly reducing computational and memory requirements, thus showing great potential in large-scale sequence modeling tasks. However, despite its breakthroughs in long-range modeling and computational efficiency, S4 still faces limitations in contextual reasoning and flexible information selection. To further enhance the model's performance in complex scenarios, the Mamba model (S6) is developed~\citep{gu2024mamba}. Mamba integrates a selection mechanism into S4, allowing the model to selectively propagate or discard information along the sequence. This improvement boosts the model's contextual reasoning abilities while maintaining the computational efficiency of S4.

Currently, Mamba has been rapidly adopted in computer vision tasks. It unfolds image patches into sequences along the horizontal and vertical dimensions of the image, and performs bi-directional scanning along these two directions. This bidirectional scanning method enables VMamba to effectively capture both global and local information within images.  At present, some studies have employed the network architectures based on visual Mamba~\citep{zhao2024rs, fu2025deep, xie2024precision, dang2024log}.  Additionally, applications in LDCT image denoising have also been reported~\citep{huangnew}. However, the classic visual Mamba scanning method (row-column scanning) tends to disrupt the original spatial continuity between adjacent pixels near image boundaries, breaking the spatial adjacency relationships, which leads to local information loss and ultimately impairs the modeling of fine structures~\citep{fu2024ssumamba}.
This separation can undermine Mamba’s performance in visual tasks, particularly in medical imaging, where capturing fine structures and lesions is critical. To address this issue, we have improved the scanning method to ensure the spatial correlation between pixels in the image. Furthermore, by combining Mamba with CNN, we harness CNN’s strengths in local feature extraction along with Mamba’s global modeling capability, allowing the model to excel in handling the complex task of LDCT image denoising.

Current researches also ignore an important issue: the inability to quantitatively characterize noise distribution in predicted images. This may lead to problems such as excessive smoothing, misjudgment of lesions and key structures, and loss of spatial resolution. The Noise Power Spectrum (NPS) can quantitatively describe the noise texture in an image~\citep{wilson2013methodology}. Therefore, this paper also  utilizes NPS to guide deep learning LDCT denoising tasks, ensuring that the predicted images more accurately restore the noise distribution present in NDCT images, thereby improving overall image quality and diagnostic value.

The contributions of this study are summarized as follows:

1. We propose CT-Mamba, a hybrid convolutional state space model designed for LDCT image denoising. This model integrates the multi-scale analysis capability of wavelet transform, the powerful local feature extraction capacity of CNN, and the long-range dependency modeling strengths of Mamba, enabling comprehensive feature capture within the images.

2. We propose the Coherence Z-Scan State Space Block (CZSS), which introduces an innovative spatially coherent Z-shaped scanning scheme. This approach ensures spatial continuity between adjacent pixels in the image, enhancing the model’s ability to preserve details and improve denoising effectiveness.

3. To ensure that the denoised LDCT images can closely restore the noise texture of NDCT, this study also designs a Mamba-driven Deep NPS loss (Deep NPS loss).

4. Using radiomics, the proposed CT-Mamba was evaluated by comparing the statistical distribution of radiomics features in different organs between denoised LDCT images and NDCT images, as well as the mean absolute error of pairwise features.

\section{Methodology}
\subsection{Preliminary: State Space Model}
The state space model (SSM) is a mathematical description of a dynamic system which maps a one-dimensional input function or sequence $x(t)\in {{\mathbb{R}}^{L}}$ to an output $y(t)\in {{\mathbb{R}}^{L}}$ through hidden states $h(t)\in {{\mathbb{R}}^{N}}$. From a mathematical perspective, this process can be represented by a linear Ordinary Differential Equation (ODE):
\begin{equation} \label{eq1}
\begin{aligned}
  h'(t) &= \mathbf{A} h(t) + \mathbf{B} x(t) \\
  y(t) &= \mathbf{C} h(t),
\end{aligned}
\end{equation}
where $\mathbf{A}\in {{\mathbb{R}}^{N\times N}}$  represents the state matrix, while $\mathbf{B}\in {{\mathbb{R}}^{N\times 1}}$ and $\mathbf{C}\in {{\mathbb{R}}^{1\times N}}$ denote the projection parameters.

In order to integrate SSM into deep learning, the linear ordinary differential equation mentioned above needs to be discretized. Typically, the discretization of the SSM uses the Zero-Order Hold (ZOH) method. By incorporating a time scale parameter $\Delta$, the continuous parameters $\mathbf{A}$ and $\mathbf{B}$ are transformed into discrete parameters $\overline{\mathbf{A}}$ and $\overline{\mathbf{B}}$, which can be defined as follows:
\begin{equation} \label{eq2}
\begin{aligned}
  & \overline{\mathbf{A}}=\exp (\Delta \mathbf{A}) \\ 
 & \overline{\mathbf{B}}={{(\Delta \mathbf{A})}^{-1}}(\exp (\Delta \mathbf{A})-I)\cdot \Delta \mathbf{B} \\ 
\end{aligned}
\end{equation}

After discretization, Equation  \eqref{eq1} can be reformulated as:
\begin{equation} \label{eq3}
\begin{aligned}
  & {h}'(t)=\overline{\mathbf{A}}h(t)+\overline{\mathbf{B}}x(t) \\ 
 & y(t)=\mathbf{C}h(t) \\ 
\end{aligned}
\end{equation}

Equation \eqref{eq3} results in a discretized state space model that can be computed recursively. However, due to its sequential nature, this discretized recursive SSM is impractical for training. To enable efficient parallelized training, this recursive process can be reformulated as a convolution for computation:
\begin{equation} \label{eq4}
\begin{aligned}
  & \overline{\mathbf{K}}=(\mathbf{C}\overline{\mathbf{B}},\mathbf{C}\overline{\mathbf{A}}\overline{\mathbf{B}},\ldots ,\mathbf{C}{{\overline{\mathbf{A}}}^{L-1}}\overline{\mathbf{B}}) \\ 
 & y=x*\overline{\mathbf{K}}, \\ 
\end{aligned}
\end{equation}
where $\overline{\mathbf{K}}\in {{\mathbb{R}}^{L}}$ represents a structured convolutional kernel, $L$ denotes the length of the input sequence $x$, and * represents the convolution operation.

The recently proposed Mamba model makes further improvements by introducing a selective scanning mechanism. This allows parameters $\mathbf{B}$, $\mathbf{C}$ and $\Delta$ to be dynamically adjusted based on the input $x$ and contextual information. As a result, Mamba can model complex temporal dynamics more effectively, as the model can adapt to the evolving characteristics of the input data.

\begin{figure*}[!t]
\centering
\centerline{\includegraphics[width=\textwidth]{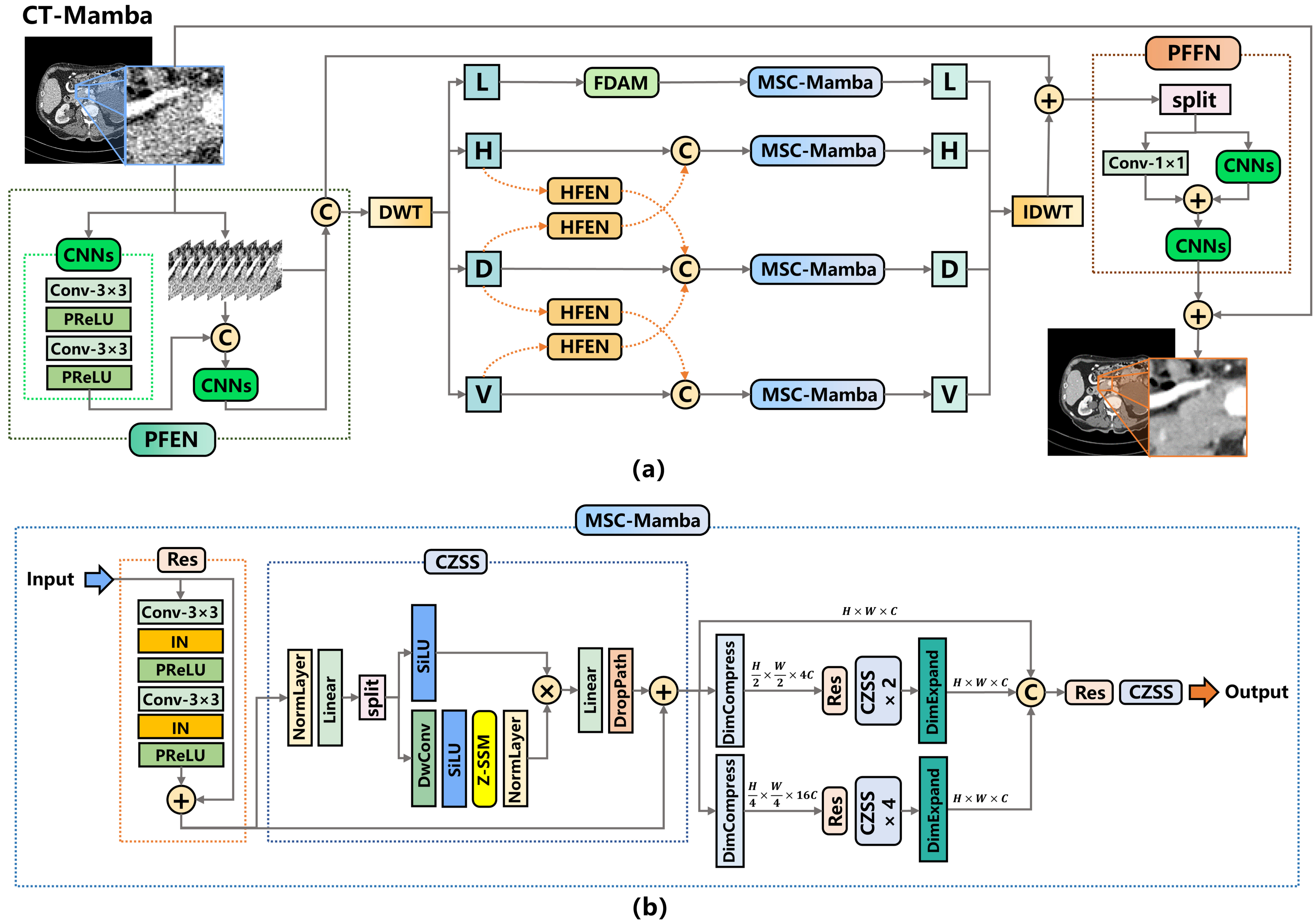}}
\caption{(a) The overall architecture of the proposed CT-Mamba. (b) The structure of the Multi-Scale Coherence  Mamba architecture (MSC-Mamba) in CT-Mamba.}
\label{fig1}
\end{figure*}

\subsection{Framework Overview}
In this paper,  we propose CT-Mamba, a hybrid convolutional state space model designed for LDCT image denoising. This model integrates the multi-scale analysis capability of wavelet transform, the powerful local feature extraction ability of CNN, and Mamba's long-range dependency modeling advantages, enabling it to comprehensively capture image features, as shown in \textcolor{blue}{Fig. \ref{fig1}(a)}.

In the initial stage of CT-Mamba, we designed a lightweight  Progressive Feature Extraction Network (PFEN). This network extracts features at two levels, gradually capturing primary spatial features of LDCT images while incorporating raw information from LDCT images to provide richer spatial information for subsequent wavelet domain processing. Then, we applied a first-level wavelet transform to decompose the spatial information into a low-frequency component, a horizontal high-frequency component, a vertical high-frequency component, and a diagonal high-frequency component. This helps CT-Mamba better capture different frequency features in LDCT images. At the end of the network, we also designed a lightweight Progressive Feature Fusion Network (PFFN), which performs feature recombination across two branches. This enables efficient integration and refinement of features extracted from the wavelet domain. These two spatial networks work in tandem, effectively capturing and processing spatial information while significantly reducing the model’s computational complexity.
 Through combined processing in both the spatial and wavelet domains, the quality of the output image is notably enhanced.

\begin{figure}[!ht]
\centerline{\includegraphics[width=0.7\columnwidth]{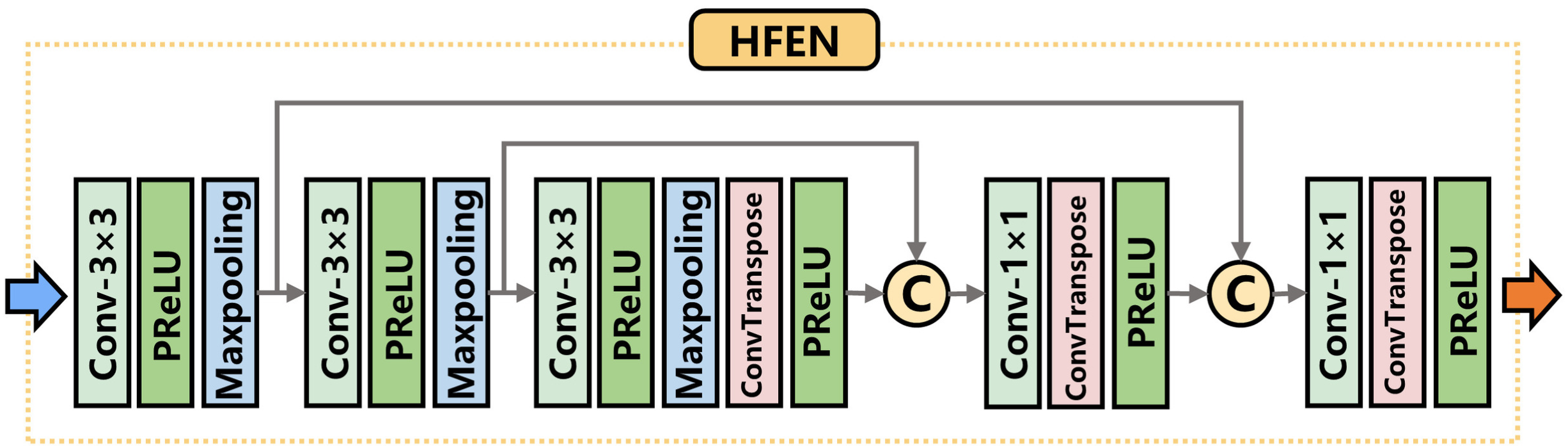}}
\caption{The structure of High-Frequency Feature Extraction Network (HFEN).}
\label{fig2}
\end{figure}

After the wavelet transform, although the components in the horizontal, vertical, and diagonal directions are decomposed independently, they still exhibit certain correlations. For example, the edges or texture details of an object may simultaneously manifest in the horizontal, vertical, and diagonal directions. We believe that the correlation between high-frequency information in the diagonal direction and that in the horizontal or vertical directions is generally stronger than the correlation between horizontal and vertical information. Based on this insight, we designed a feature fusion strategy incorporating a High-Frequency Feature Extraction Network (HFEN), as shown in \textcolor{blue}{Fig.} \ref{fig2}. The HFEN is equipped with multi-scale feature extraction, cross-scale feature fusion, and channel compression capabilities, which significantly enhances its ability to extract features from frequency information in different directions. By fusing high-frequency information from the horizontal and vertical directions into the diagonal direction, and vice versa, this strategy facilitates effective interactions among high-frequency components. This enables the network to better capture complex structural features within the image, thereby improving overall performance.

\begin{figure}[!ht]
\centerline{\includegraphics[width=0.7\columnwidth]{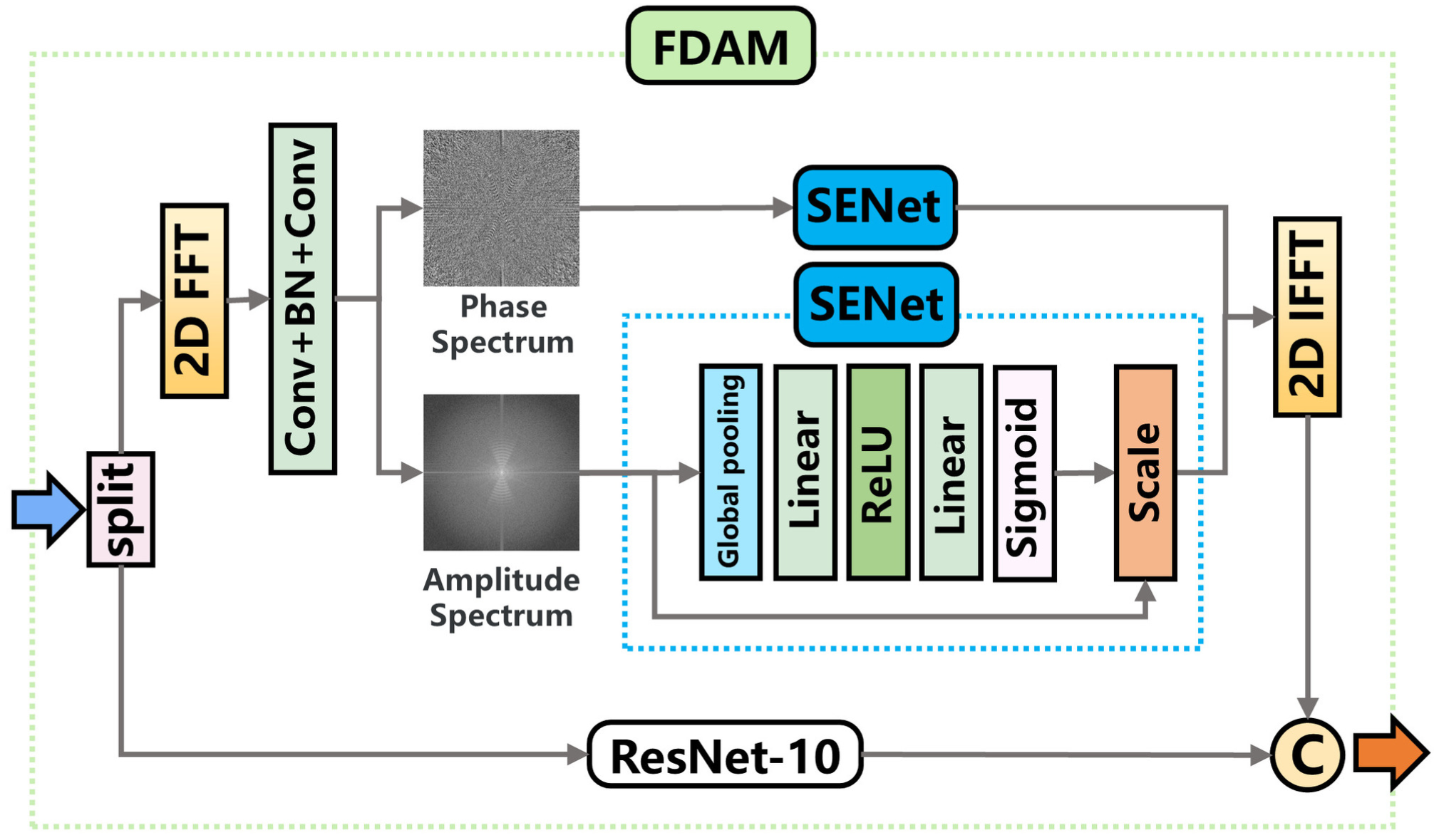}}
\caption{The structure of Frequency Domain Attention Module (FDAM).}
\label{fig4}
\end{figure}

We designed a Multi-Scale Coherence Mamba architecture (MSC-Mamba) to learn the features of each frequency branch obtained by wavelet transform decomposition, as shown in \textcolor{blue}{Fig. \ref{fig1}(b)}.  MSC-Mamba captures and fuses frequency features learned from three different scales ($H\times W\times C$, $\frac{H}{2}\times \frac{W}{2}\times 4C$, and $\frac{H}{4}\times \frac{W}{4}\times 16C$) by integrating Coherence Z-Scan State Space Block (CZSS). At lower scales, MSC-Mamba effectively captures global feature information, while at higher scales, it focuses more on local details. By combining feature learning across different scales, MSC-Mamba effectively  ensures spatial coherence, providing superior denoising performance. Additionally, inspired by \citep{guo2023spatial}, we designed a Frequency Domain Attention Module (FDAM) based on Fourier transform to enhance the low-frequency features in the wavelet domain, as shown in \textcolor{blue}{Fig.} \ref{fig4}. Each grid in the extracted Fourier spectrum contains global information on the low-frequency features, enabling efficient modeling of long-range spatial dependencies in the frequency domain. Enhanced by FDAM, richer low-frequency characteristics are provided for MSC-Mamba located in the low-frequency branch, thereby optimizing the modeling of overall structural information and improving CT-Mamba’s performance.


\subsection{Coherence Z-Scan State Space Block}
\begin{figure*}[!ht]
\centerline{\includegraphics[width=\textwidth]{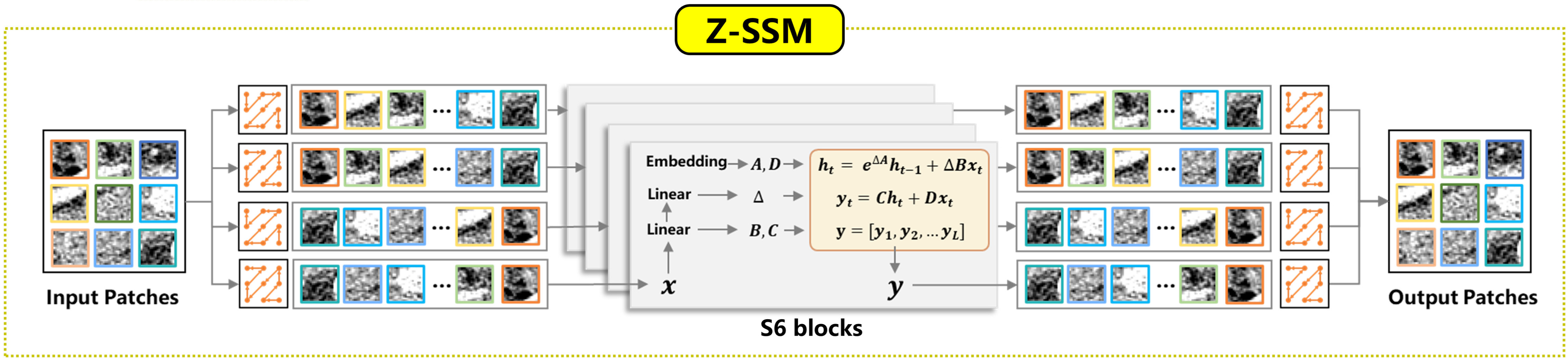}}
\caption{Illustration of the structure of Z-SSM. The Z-shaped scanning method is adopted to ensure spatial continuity between each adjacent pixel in the image.}
\label{fig5}
\end{figure*}

The Coherence Z-Scan State Space Block (CZSS) is a novel feature extraction unit specifically designed for tasks like medical imaging. Classic visual Mamba models, such as VMamba, which excel in natural image processing, utilize a row-column scanning order with bi-directional scanning in both horizontal and vertical directions. However, such scanning methods can overly separate some adjacent pixels when unfolding an image into a sequence. For instance, in horizontal scanning, there is a substantial distance between the end of one row and the beginning of the next,  weakening spatial connections and subsequently hindering the capture of fine structures. To address this issue, inspired by the Zigzag scanning method in JPEG image compression theory, Z-SSM adopts a similar Z-shaped scanning approach to maintain spatial continuity between each adjacent pixel in the image, as shown in \textcolor{blue}{Fig.} \ref{fig5}. Z-SSM employs multi-directional scanning and combined with a selective state space model SSM (S6) for modeling, ultimately merging the sequence and restoring it to a two-dimensional structure. This design ensures that subtle lesions and detailed information within the image are not overlooked, thereby enhancing the denoising effect of LDCT images.

A residual block with two convolutional layers is added to the front of CZSS can locally maintain the frequency-spatial correlation of input features and efficiently transfer local contextual information, thereby enabling more effective utilization and further optimization of these features, as represented by ``Res" in \textcolor{blue}{Fig. \ref{fig1}(b)}. The CZSS begins by standardizing the input data through layer normalization, improving training stability. Next, a linear transformation adjusts the data dimensions, followed by depthwise separable convolution, which operates on each input channel independently, reducing parameter count and focusing on feature extraction. After convolution, the SiLU activation function is applied, and its smooth gradient properties help the model to stably learn nonlinear features, thereby enhancing its ability to capture complex patterns. Then, Z-SSM performs Z-shaped scanning across four paths, after which the outputs are merged and restructured into a 2D form. Layer normalization is then applied, and the SiLU activation  is used to adjust deep features.  Finally, a linear layer is applied with DropPath to enhance robustness and mitigate overfitting. CZSS finds a good balance between computational efficiency and feature extraction capability, allowing for stacking more CZSS blocks under similar depth constraints.

\begin{figure*}[!t]
\centerline{\includegraphics[width=\textwidth]{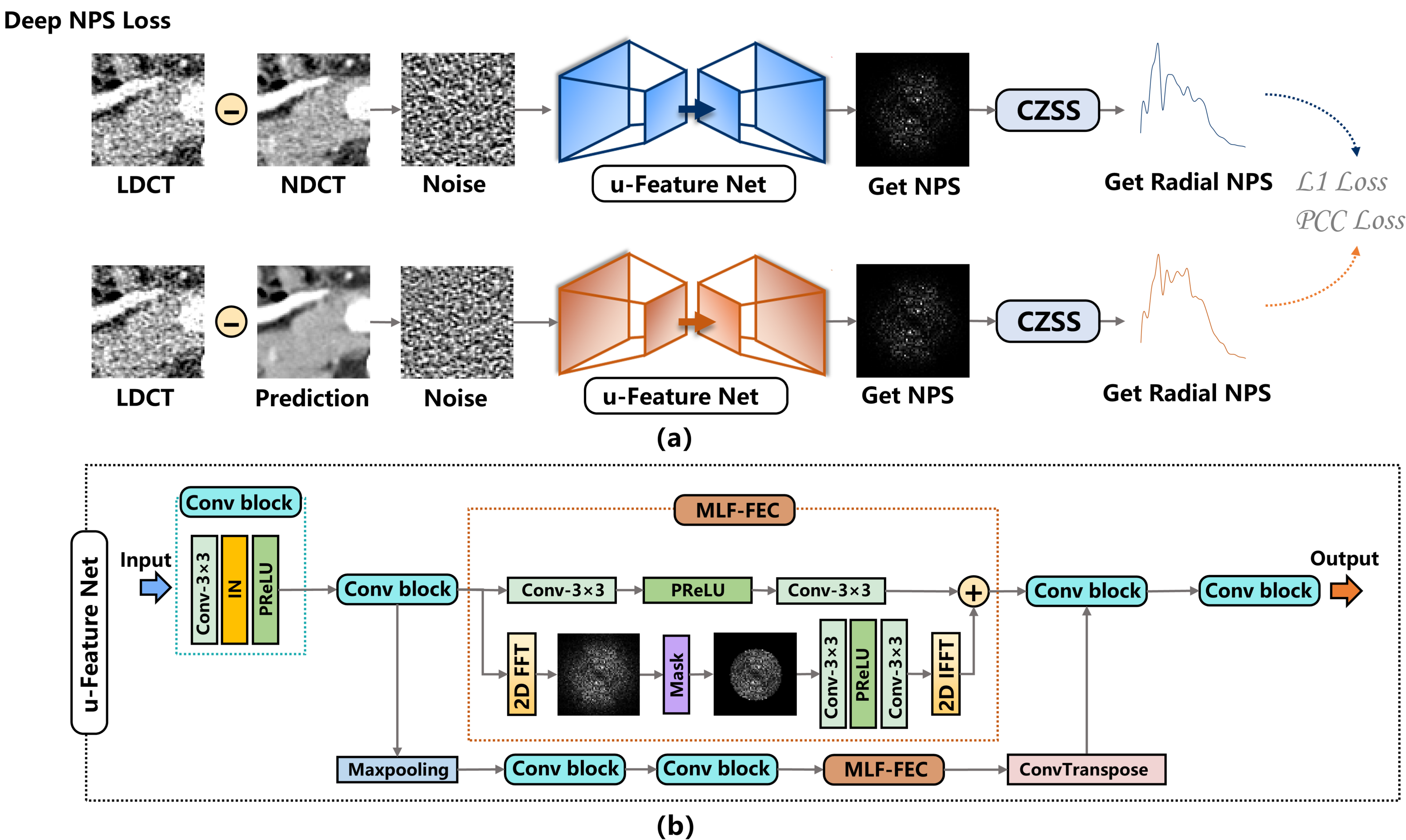}}
\caption{Detailed structure of the dual-branch Deep NPS loss and its key components. (a) Structure of the Deep NPS loss. (b) Detailed structure of the u-Feature Net.}
\label{fig6}
\end{figure*}

\subsection{Deep Noise Power Spectrum Loss}
Noise texture is highly sensitive to changes in subjective visual quality. NPS, based on Fourier transform, can characterize the correlation between pixels from a frequency domain perspective, thereby providing a quantitative description of image noise texture. In CT imaging, noise originates from various factors, including but not limited to the signal acquisition system, radiation dose, and computational algorithms.

In LDCT denoising tasks, neglecting the optimization of noise texture can severely impact the quality of denoised images. Therefore, quantitative analysis of the NPS is crucial for optimizing and evaluating image quality. In this study, we utilize NPS to guide the denoising of LDCT images for the first time, aiming to more accurately restore the noise texture in the denoised images. The NPS in this study is defined as follows:
\begin{equation} \label{eq5}
\text{NP}{{\text{S}}_{2D}}(u,v)=\frac{{{p}_{x}}{{p}_{y}}}{{{N}_{x}}{{N}_{y}}}{{\left| \text{DFT}(\text{noise(}m,n)) \right|}^{2}},
\end{equation}
where ${{p}_{x}}$ and ${{p}_{y}}$ represent the sampling intervals, ${{N}_{x}}$ and ${{N}_{y}}$ denote the two dimensions of the region of interest (ROI), and $\text{noise(}m,n)$ is the noise image obtained by subtracting two images. $\text{(}m,n) $ and $ (u,v) $ represent coordinates in the spatial and frequency domains, respectively.

Since it is difficult for medical images to separate noise from a single image, we designed a dual-branch structure optimized based on NPS, as shown in \textcolor{blue}{Fig. \ref{fig6}(a)}. By subtracting the NDCT image from the LDCT image, we obtain reference noise (denoted as $\text{nois}{{\text{e}}_{\text{GT}}}$, $\text{nois}{{\text{e}}_{\text{GT}}}={{I}_{\text{LDCT}}}-{{I}_{\text{NDCT}}}$), and by subtracting the model-predicted image from the LDCT image, we obtain the predicted noise (denoted as $\text{nois}{{\text{e}}_{\text{Pred}}}$, $\text{nois}{{\text{e}}_{\text{Pred}}}={{I}_{\text{LDCT}}}-{{I}_{\text{Pred}}}$). As each image can be abstracted as a combination of signal and noise, so the difference between these two noise components lies solely in the noise present in the model-predicted image versus the NDCT image. By analyzing NPS based on the pair of noise, we can guide the model to generate images with a noise distribution closer to that of NDCT images.

In the dual-branch structure, we designed an enhanced U-shaped network (u-Feature Net) to extract features from $\text{nois}{{\text{e}}_{\text{GT}}}$ and $\text{nois}{{\text{e}}_{\text{Pred}}}$, as shown in \textcolor{blue}{Fig. \ref{fig6}(b)}. Low-frequency information is crucial in noise modeling, particularly for maintaining the stability of the overall noise structure and texture.  Effective capture of low-frequency features is thus essential for improving noise texture. To address this, we designed  a Mid and Low Frequency Feature Enhancement Connection (MLF-FEC) within u-Feature Net, specifically reinforcing the network’s processing of mid- and low-frequency information. Building on this, we applied an NPS transformation to the extracted noise features and incorporated the long-range modeling capability of CZSS to accurately capture the global noise feature distribution across the full frequency. This design effectively guides CT-Mamba in optimizing noise distribution.

\begin{figure}[!ht]
\centerline{\includegraphics[width=0.3\columnwidth]{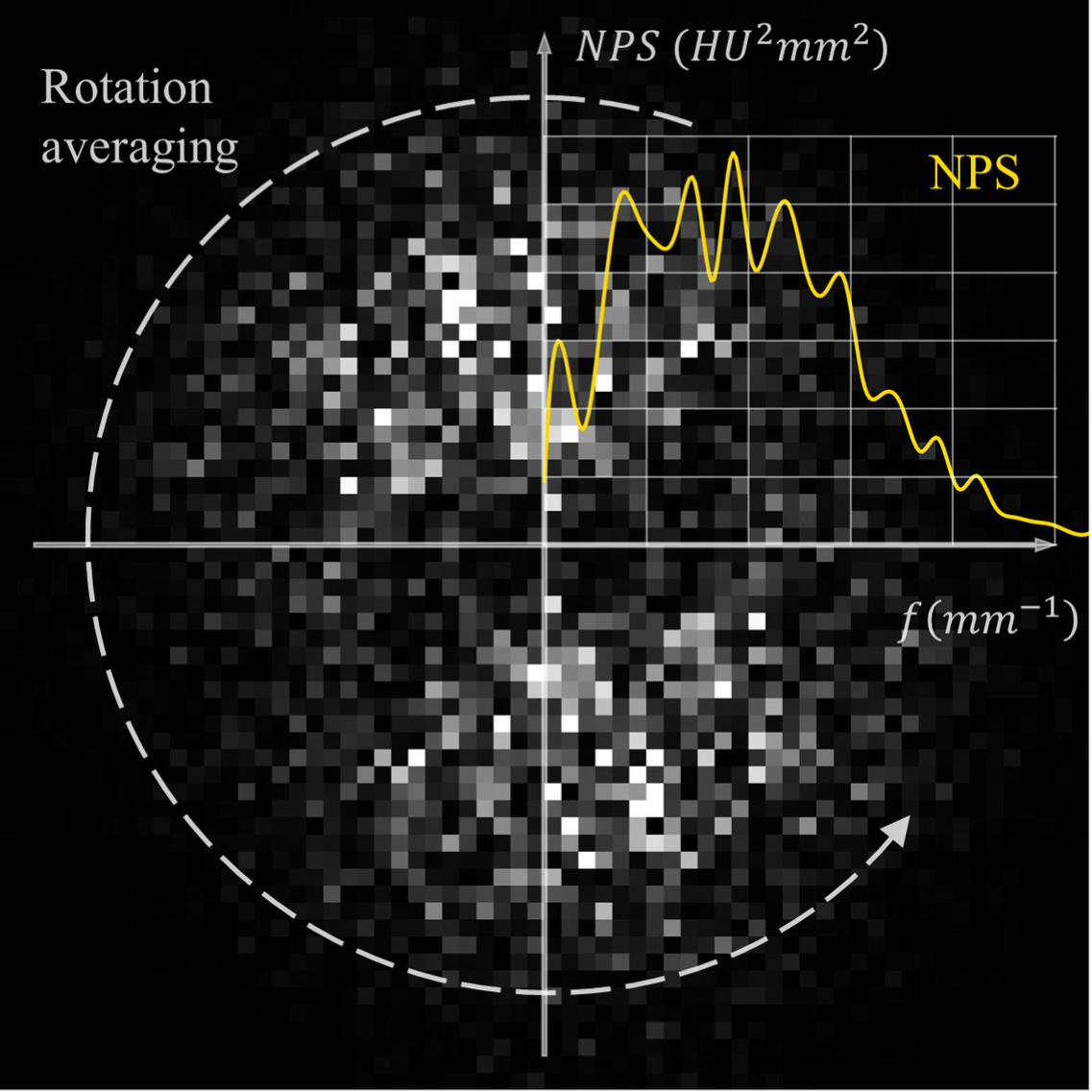}}
\caption{Schematic diagram of the radial transformation process of a 2D NPS signal.}
\label{fig7}
\end{figure}

After the full-band modeling of CZSS, the extracted $\text{nois}{{\text{e}}_{\text{GT}}}$ and $\text{nois}{{\text{e}}_{\text{Pred}}}$ features are converted into one-dimensional radial NPS (Radial NPS) to reduce complexity while retaining critical frequency information. The radialization principle is shown in \textcolor{blue}{Fig.} \ref{fig7}. We then use L1 loss to measure the difference between the two radial NPS signals and design a correlation loss based on the Pearson correlation coefficient to guide model optimization. This approach ensures that the noise texture in the model’s denoised images more closely resembles that of NDCT images, thereby improving overall image quality and diagnostic value.

\subsection{Overall Objective}
The loss function measures the difference between the model's predicted output (the denoised LDCT image) and the ground truth (NDCT image). While the model architecture determines the model's complexity, the loss function controls how denoising characteristics are learned from the training dataset. To recover high-quality denoised CT images from LDCT, we combine different loss components into a hybrid objective function:

1) Pixel-level L1 Loss: Compared with L2 loss (mean square error, MSE), L1 loss does not overly penalize large errors between the predicted and NDCT images,  so it can alleviate the blurriness and unnatural effects caused by L2 loss to a certain extent. In our LDCT denoising task, it is defined as~\eqref{eq6}:
\begin{equation} \label{eq6}
{{\mathcal{L}}_{\text{L}1}}(y,\hat{y})={{\left\| \overset{\wedge }{\mathop{y}}\,-y \right\|}_{1}},
\end{equation}
where $\overset{\wedge }{\mathop{y}}\,$ represents the model's predicted output, $y$ represents the ground truth NDCT image.

2) Deep NPS loss: This loss component ensures that the noise texture in the denoised image predicted by CT-Mamba is closer to that of the NDCT image. It is represented in conjunction with the description in Section 2.4 as~\eqref{eq7}:
\begin{equation} \label{eq7}
\begin{aligned}
{{\mathcal{L}}_{\text{deep-NPS}}}(x,y,\hat{y}) = & \, {{\gamma }_{1}} \cdot {{\left\| \text{NP}{{\text{S}}_{\text{radial}}}(\phi (x,y)-\phi (x,\hat{y})) \right\|}_{1}} \\
& \hspace{-8em} + {{\gamma }_{2}} \cdot \left( 1 - \rho \left( \text{NP}{{\text{S}}_{\text{radial}}}(\phi (x,y)), \text{NP}{{\text{S}}_{\text{radial}}}(\phi (x,\hat{y})) \right) \right),
\end{aligned}
\end{equation}
where $x$ represents the input LDCT image, ${{\gamma }_{1}}$ and ${{\gamma }_{2}}$ are weighting hyperparameters that respectively control the contributions of the L1 loss term and the correlation loss term. $\phi (x,y)$ and $\phi (x,\hat{y})$ represent the noise features extracted from the input image $x$ and the real image $y$, and the input image $x$ and the predicted image $\overset{\wedge }{\mathop{y}}\,$ through a series of networks in each branch. $\text{NP}{{\text{S}}_{\text{radial}}}(\cdot )$ represents the process of obtaining the radial NPS, and $\rho (\cdot ,\cdot )$ indicates the Pearson correlation coefficient, which measures the correlation between the two radial NPS, aiming to preserve the structural consistency of the noise texture.

3) Feature-Level Perceptual Loss: Perceptual loss captures information at the feature level using a pre-trained deep network.  This loss is based on ResNet50~\citep{he2016deep}, pre-trained on the ImageNet dataset, with its weights frozen during training. Since the pre-trained ResNet50 accepts color images as input, we replicated the grayscale CT images across three channels to fit this input requirement. This approach has been validated in LDCT tasks~\citep{yang2018low}. we extracted feature maps after its four main stages and used L2 loss to measure the similarity between the CT-Mamba output and the ground truth image.  The perceptual loss is defined as~\eqref{eq8}:
\begin{equation} \label{eq8}
\mathcal{L}_{\text{perceptual}}(y, \hat{y}) = \sum_{i=1}^{4} \| f_i(y) - f_i(\hat{y}) \|_2
,
\end{equation}
where ${{f}_{i}}$ represents the feature map extracted at stage $i$, $y$ and $\overset{\wedge }{\mathop{y}}\,$ denote the ground truth and predicted images, respectively.

In summary, the overall objective function is the weighted sum of each loss term~\eqref{eq9}: 
\begin{equation} \label{eq9}
{{\mathcal{L}}_{\text{total}}}={{\lambda }_{1}}\cdot {{\mathcal{L}}_{\text{L1}}}+{{\lambda }_{2}}\cdot {{\mathcal{L}}_{\text{deep-NPS}}}+{{\lambda }_{3}}\cdot {{\mathcal{L}}_{\text{perceptual}}},
\end{equation}
where ${{\lambda }_{2}}=({{\gamma }_{1}},{{\gamma }_{2}})$.

\section{Experiment designs and results}
\subsection{Datasets}
In this work, we used a publicly released patient dataset for the 2016 NIH-AAPM-Mayo Clinic Low-Dose CT Grand Challenge (referred to as the Mayo dataset)~\citep{mccollough2017low}. This dataset contains paired low-dose and normal-dose abdominal images with a slice thickness of 1 mm for 10 patients.  The quarter-dose LDCT images were generated by adding Poisson noise to the projection data of the NDCT images to mimic a noise level that corresponded to 25\% of the full dose. To ensure the statistical robustness of our quantitative evaluations, we implemented a five-fold cross-validation strategy on the Mayo dataset. Specifically, the 10 patients were divided into 5 groups, with each fold using 2 patients as the test and the remaining 8 for training. For the visual analysis, we used data from 9 patients for model training, totaling 5,410 image pairs, while data from the remaining  patient (L506) was used for testing model performance, totaling 526 pairs. Additionally, using the method  in~\citep{wang2022simulating},  we simulated paired low-dose and normal-dose images for 20 patients from the AMOS dataset (referred to as the Simulator dataset) for model training and validation~\citep{ji2022amos}. The low-dose images also correspond to a quarter of the normal dose. We used data from 19 patients for training, totaling 1,872 image pairs, and data from 1 patient (AMOS-0033) for testing model performance, totaling 106 pairs.

We also used in-house real low- and normal-dose head phantom data for experimental validation. Specifically, an anthropomorphic head phantom was scanned using a Cone-Beam Computed Tomography (CBCT) on-board imager (TrueBeam System, Varian Medical Systems, Palo Alto, CA). The phantom data comprises pairs of CT images acquired using low-dose parameters (80 kV, 100 mA) and normal-dose parameters (80 kV, 400 mA), with the low-dose settings representing one-quarter of the radiation exposure of the normal-dose configurations. The data directly reflect the noise and artifact characteristics under real low- and normal-dose conditions in daily CBCT for patient setup in image-guided radiotherapy.

\subsection{Implementation details}
The proposed CT-Mamba model is implemented in PyTorch and trained in an environment equipped with an NVIDIA RTX 3090 Ti 24G GPU. The optimizer used is AdamW, with parameters set to ($\beta 1$, $\beta 2$) = (0.9, 0.999) and a weight decay of 0.02. The initial learning rate is set to 1e-3 and is gradually reduced to 1e-6 using cosine annealing. During training, each image is randomly cropped into four 64×64 patches as input for each batch, with the batch size of 8. To balance the numerical relationship between the Deep NPS loss and the L1 loss and ensure training stability, the Deep NPS loss is not introduced during the first 10 epochs. Afterward, the hyperparameters of corresponding weights for the Deep NPS loss are set to ${{\lambda }_{2}}$ = (1e-4, 1e-2). The perceptual loss weight ${{\lambda }_{3}}$ is set to 1e-2. The training process lasts for 250 epochs, with the final model selected based on the optimal loss achieved.

\subsection{Experiment result and design}
To demonstrate the effectiveness of the proposed CT-Mamba, we selected several advanced representative methods for comparative experiments, including RED-CNN~\citep{chen2017low}, Uformer~\citep{wang2022uformer}, CTformer~\citep{wang2023ctformer}, DDPM, and VM-UNet~\citep{ruan2024vm}. RED-CNN is CNN-based methods, Uformer and CTformer are Transformer-based methods, DDPM is a diffusion model-based method, and VM-UNet is based on the Mamba framework.

\begin{figure}[!ht]
\centerline{\includegraphics[width=0.8\columnwidth]{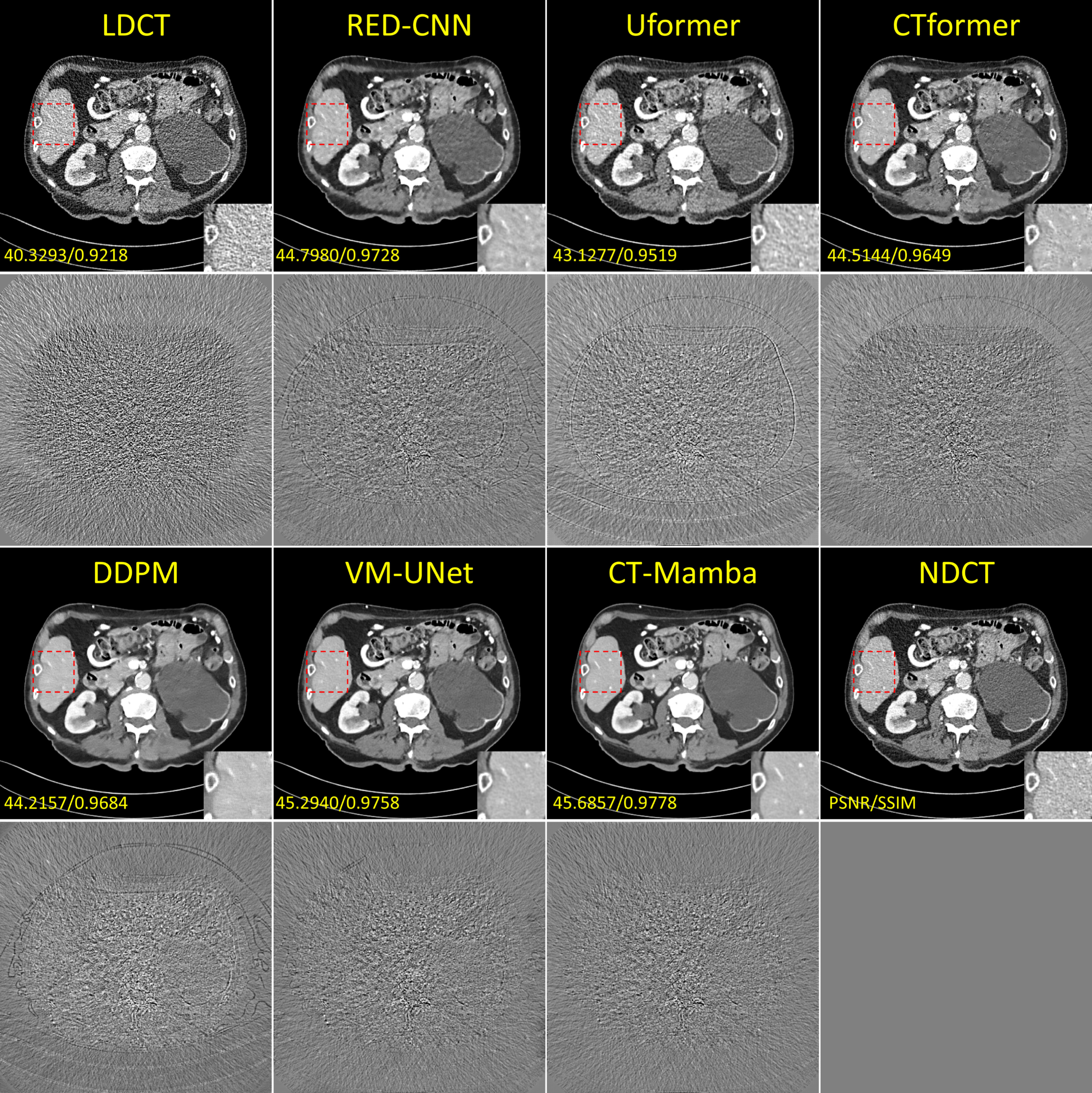}}
\caption{A set of slice prediction results and difference images obtained from the Mayo dataset L506 patient. The enlarged ROI (blood vessel) marked by the red dashed rectangle is located in the lower right corner of the image.}
\label{fig8}
\end{figure}

\subsubsection{Visual evaluation}
\begin{figure}[!ht]
\centerline{\includegraphics[width=0.8\columnwidth]{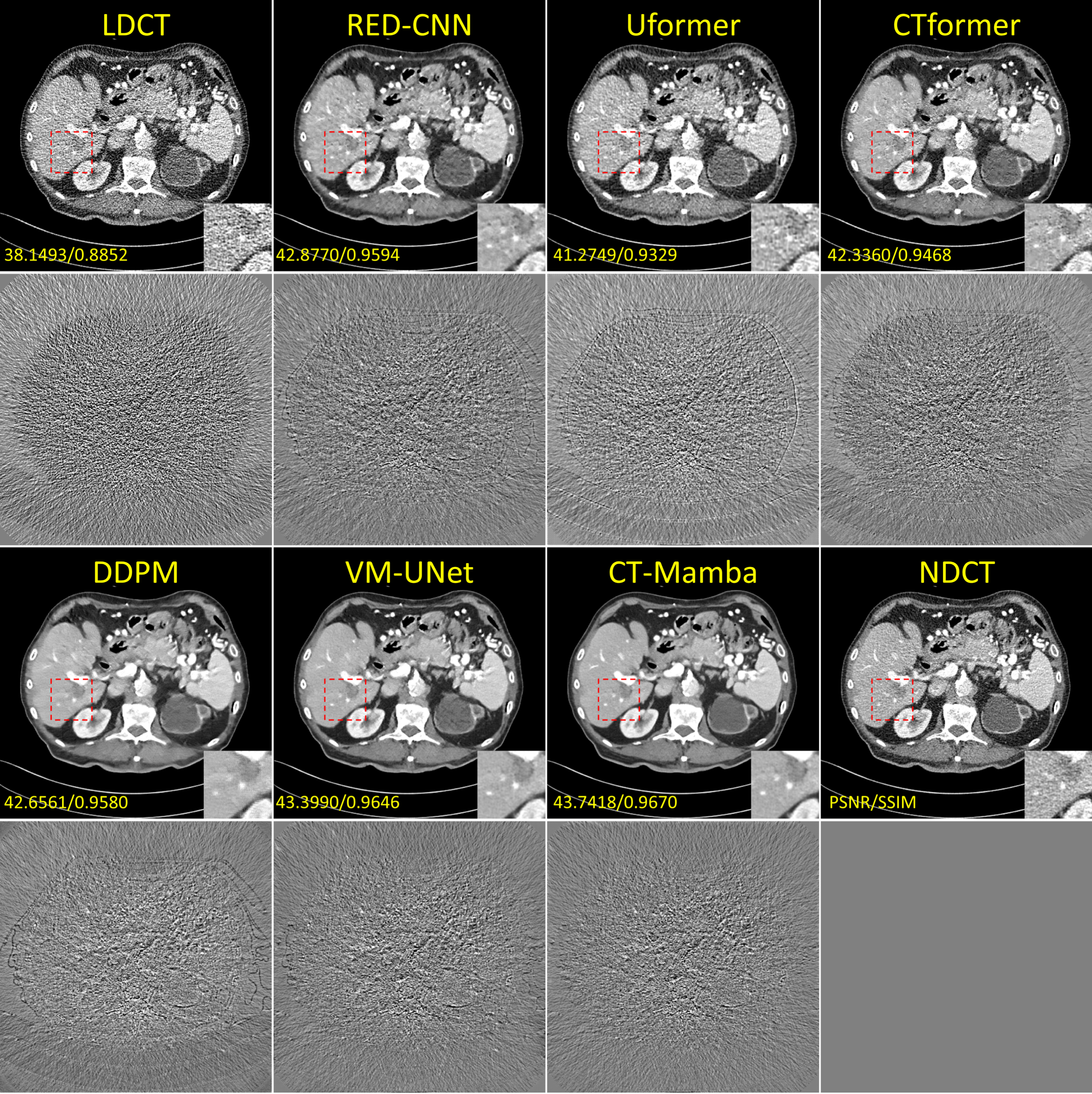}}
\caption{A set of slice prediction results and difference images obtained from the Mayo dataset L506 patient. The enlarged ROI (low-attenuation lesion, tissue edge) marked by the red dashed rectangle is located in the lower right corner of the image.}
\label{fig9}
\end{figure}

This section presents the visual validation results of the CT-Mamba model compared to various baseline methods. In the Mayo dataset, we selected two representative slices, each containing the predicted results from each method along with their difference images relative to the NDCT images, to validate  the comprehensive performance of our proposed method in LDCT processing, as shown in \textcolor{blue}{Fig.} \ref{fig8} and \textcolor{blue}{Fig.}  \ref{fig9}. To further assess detail representation, an enlarged view of the ROI which is marked by a red dashed rectangle  is provided in the lower right corner of each image, along with the PSNR/SSIM quantitative results for each method displayed in the lower left corner.
From the results shown in \textcolor{blue}{Fig.} \ref{fig8} and \textcolor{blue}{Fig.} \ref{fig9}, all methods exhibit a certain degree of noise and artifact removal capability. However, compared to other methods, our proposed approach achieves the optimal processing performance, demonstrating excellent denoising competence in terms of detail preservation and artifact suppression. In the ROIs, we observe that CT-Mamba enhances the representation of fine structures (such as blood vessels), as shown in \textcolor{blue}{Fig.} \ref{fig8}; low-attenuation lesions are more clearly visible while maintaining well-defined tissue edges, as shown in \textcolor{blue}{Fig.} \ref{fig9}.

\begin{figure}[!ht]
\centerline{\includegraphics[width=0.8\columnwidth]{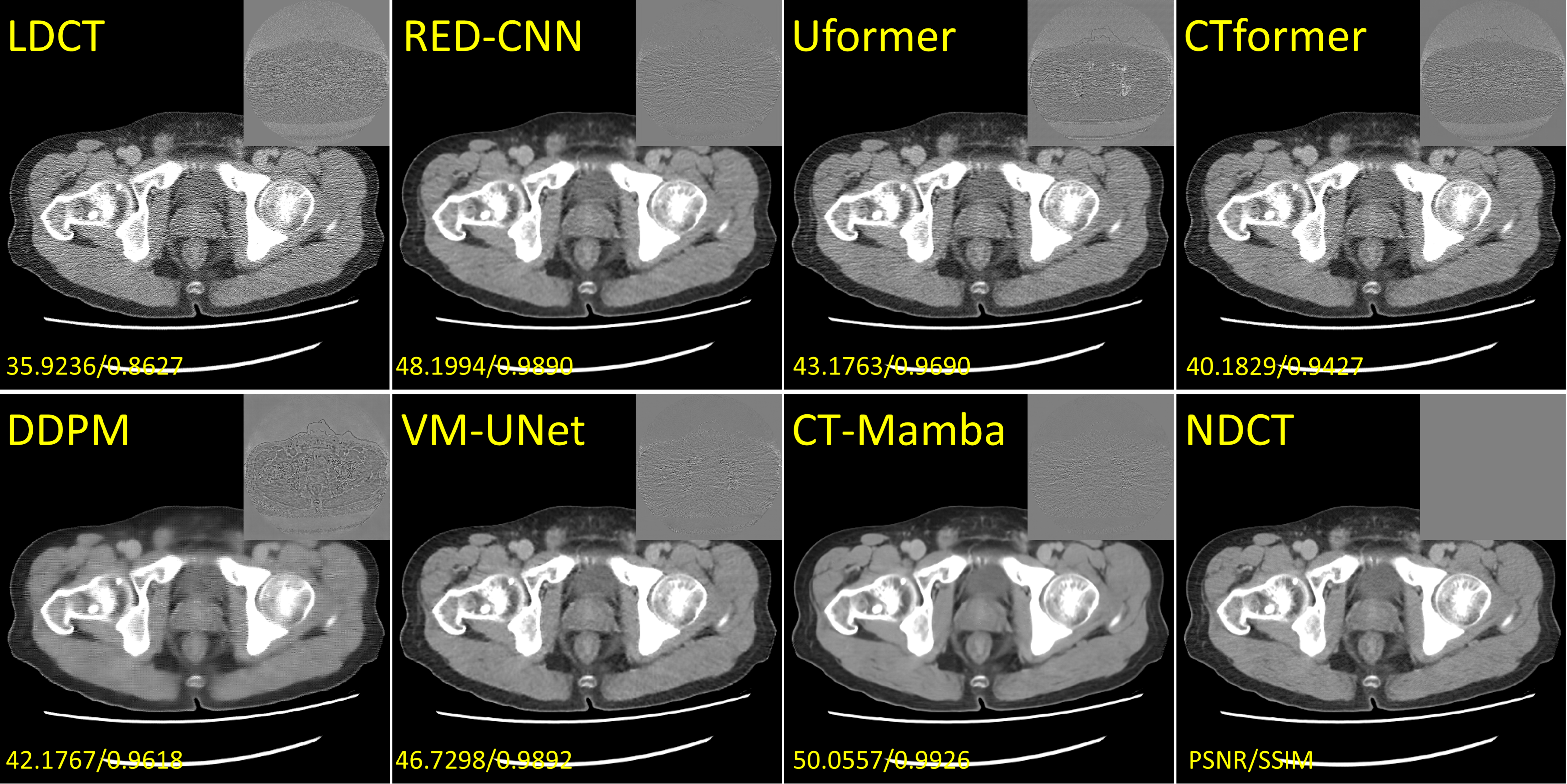}}
\caption{A set of slice prediction results and difference images obtained from the Simulator dataset AMOS-0033 patient, with the difference images located in the upper right corner.}
\label{fig10}
\end{figure}

\begin{figure}[!ht]
\centerline{\includegraphics[width=0.8\columnwidth]{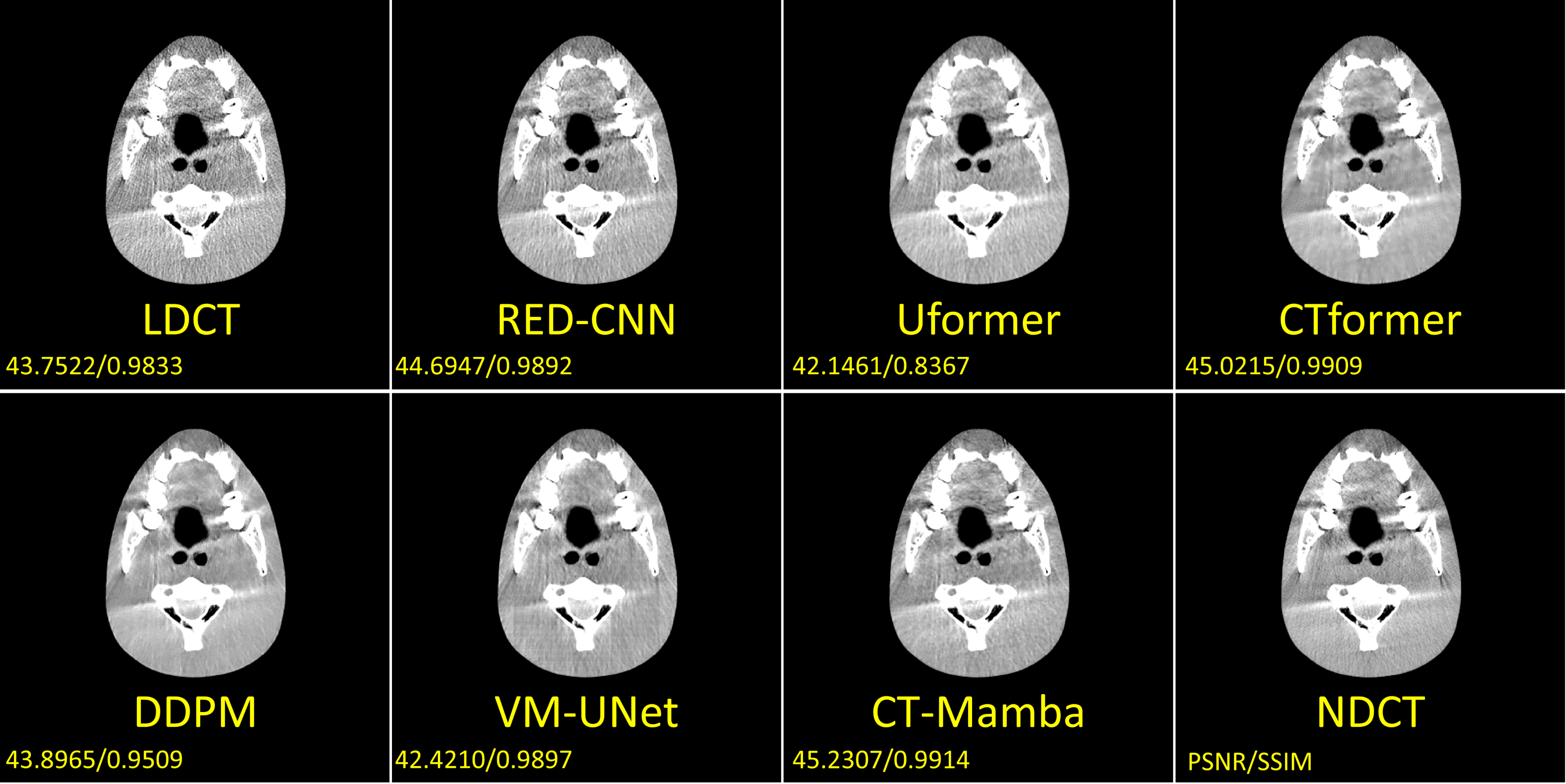}}
\caption{Visual comparison results on the real low- and normal-dose   head phantom data, which comprises pairs of CT images acquired using on-board CBCT scanner with low-dose (80 kV, 100 mA) and normal-dose (80 kV, 400 mA) settings, respectively. 
}
\label{phantom}
\end{figure}

In \textcolor{blue}{Fig.} \ref{fig10}, we also present the denoised results and difference images of a representative slice from the Simulator dataset, further validating the outstanding denoising performance and generalization capability of CT-Mamba across different datasets. The upper right corner of each predicted image shows the corresponding difference image. Additionally, we selected one representative case from the real dose head phantom data, as shown in \textcolor{blue}{Fig.} \ref{phantom}. CT-Mamba effectively reduces noise while achieving the best perceptual quality, further demonstrating the generalizability of our method in clinical scenarios.

For the Mayo and Simulator datasets, we set the window level of the predicted results to 40 HU and the window width to 400 HU (i.e., a range from -160 HU to 240 HU). For the difference images, the window level is set to 0 HU with a window width of 200 HU (i.e., a range from -100 HU to 100 HU). For the real dose head phantom data, the window level was set to 0 HU and the window width to 800 HU (i.e., a range from -400 HU to 400 HU).

\begin{table}
\small
\caption{Quantitative comparison of various methods on the Mayo dataset under five-fold cross-validation.}
\centering 
\setlength{\tabcolsep}{5pt}
\renewcommand{\arraystretch}{0.9} 
\begin{tabular}{c c c c c}
\hline
Method   & PSNR $\uparrow$ & SSIM $\uparrow$ & RMSE $\downarrow$ & VIF $\uparrow$\\
\hline
LDCT     & 40.8926±0.8687  & 0.9309±0.0113   & 37.8652±3.7472    & 0.2709±0.0150 \\
RED-CNN  & 44.8327±1.6669  & 0.9714±0.0144   & 24.2221±5.2109    & 0.3316±0.0231 \\
Uformer  & 46.1255±0.7708  & 0.9802±0.0032   & 20.5797±1.8107    & 0.3557±0.0167 \\
CTformer & 44.5024±0.7586  & 0.9710±0.0045   & 24.8519±2.1773    & 0.3100±0.0144 \\
DDPM     & 42.8780±1.2888  & 0.9533±0.0143   & 29.8810±4.7620    & 0.3020±0.0151 \\
VM-Unet  & 45.4553±0.8614  & 0.9784±0.0035   & 22.4240±2.4696    & 0.3417±0.0169 \\
\textbf{CT-Mamba}&\textbf{46.2251±0.8006}&\textbf{0.9805±0.0033}&\textbf{20.3479±1.8592
}&\textbf{0.3587±0.0180}\\
\hline
\end{tabular}
\label{table1} 
\end{table}

\subsubsection{Quantitative analysis}
Our method not only demonstrates significant optimization effects in subjective visual evaluation but also achieves excellent performance across multiple quantitative metrics. Most studies use common quantitative metrics in image processing to report their results, such as Peak Signal-to-Noise Ratio (PSNR), Structural Similarity Index Measure(SSIM), and Root Mean Square Error (RMSE). However, existing studies have indicated that these metrics do not sufficiently reflect  clinical relevance~\citep{patwari2023reducing}. Although human readers are considered as the gold standard for evaluating medical images, conducting multi-reader studies is time-consuming and costly. The research by Eulig et al. indicates that, compared to quantitative metrics like SSIM and PSNR for CT and MR images, various metrics including Visual Information Fidelity (VIF) have a higher correlation with human ratings~\citep{eulig2024benchmarking}. Therefore, this study employs PSNR, SSIM, RMSE, and VIF for comprehensive evaluation. Additionally, in next section, we conducted a radiomics analysis to extract and compare the similarity of radiomic features.

\begin{table}
\small
\caption{The average quantitative results of different methods for patient (AMOS-0033) in the Simulator test dataset.}
\centering 
\setlength{\tabcolsep}{5pt}
\renewcommand{\arraystretch}{0.9} 
\begin{tabular}{c c c c c}
\hline
Method   & PSNR $\uparrow$ & SSIM $\uparrow$ & RMSE $\downarrow$ & VIF $\uparrow$\\
\hline
LDCT     & 38.7673        & 0.9099         & 35.5206          & 0.3573       \\
RED-CNN  & 49.7609        & 0.9921         & 9.8209           & 0.5098       \\
Uformer  & 44.9416        & 0.9795         & 17.1556          & 0.4241       \\
CTformer & 43.1905        & 0.9673         & 21.4047          & 0.4063       \\
DDPM     & 43.7161        & 0.9696         & 19.7144          & 0.3776       \\
VM-Unet  & 48.0132        & 0.9917         & 11.9833          & 0.4880       \\
\textbf{CT-Mamba}&\textbf{51.2054}&\textbf{0.9942}&\textbf{8.2936}&\textbf{0.5444}\\
\hline
\end{tabular}
\label{table2} 
\end{table}

\begin{table}
\small
\caption{The average PSNR, SSIM, RMSE, and VIF results of the real dose head phantom data.}
\centering 
\setlength{\tabcolsep}{5pt}
\renewcommand{\arraystretch}{0.9} 
\begin{tabular}{c c c c c}
\hline
Method   & PSNR $\uparrow$ & SSIM $\uparrow$ & RMSE $\downarrow$ & VIF $\uparrow$\\
\hline
LDCT     & 42.7276       &  0.9703        &   57.0131      &   0.2719     \\
RED-CNN  &  44.0203      &  0.9809        &   48.9109      &   0.2880     \\
Uformer  &  42.1097    &  0.8435        &   60.1177      &   0.2756    \\
CTformer &   44.6989     &  \textbf{0.9860}        &   45.1705     &    0.2862    \\
DDPM     &   43.7502     &   0.9408      &   49.9756      &      0.2859        \\
VM-Unet  &   44.0641      &   0.9857       &   48.3067        &   0.2847     \\
\textbf{CT-Mamba}&\textbf{44.9698}&0.9859&\textbf{43.7079}&\textbf{0.2963}\\
\hline
\end{tabular}
\label{table_phantom} 
\end{table}

In Table \ref{table1} and Table \ref{table2}, we present the average quantitative results of PSNR, SSIM, RMSE, and VIF obtained from five-fold cross-validation on the Mayo dataset and from patient AMOS-0033 in the Simulator dataset, respectively.  A comprehensive evaluation of the PSNR, SSIM, RMSE, and VIF metrics shows that CT-Mamba model outperforms the comparison methods across all quantitative metrics.
Similarly, in Table \ref{table_phantom}, we present the average quantitative results of the real dose head phantom data. The results demonstrate that CT-Mamba maintains excellent performance under real-dose conditions, further demonstrating its potential for clinical application.

\subsubsection{Radiomics reserach}
Radiomics holds the potential to transform digital medical images into quantitative features that reveal underlying pathology, with promising applications in tumor classification and patient outcome prediction~\citep{xia2023predicting}. In this study, the TotalSegmentator~\citep{wasserthal2023totalsegmentator} extension in 3D Slicer with a pretrained nnU-Net~\citep{isensee2021nnu} was utilized for automatic segmentation of target organs (aorta, right kidney, liver, stomach and small bowel), followed by manual adjustments to ensure accuracy. Under the five-fold cross-validation, for all patients in the Mayo dataset, 167 radiomics features (including 18 first-order features, 75 texture features and 74 wavelet-HHH features) were extracted from the volumes of each target organ using PyRadiomics (ver. 3.0). An ideal LDCT denoising algorithm is expected to produce  denoised volumes with statistically similar radiomics features to those of NDCT volumes. The following two tests were designed to show statistical similarity:

(1) \textbf{Statistical Distribution of First-Order Features:} In statistical analysis, the Wilcoxon rank-sum test is commonly used to assess whether two datasets share the same distribution. For each volume, the Wilcoxon rank-sum test was employed to evaluate the consistency of radiomics features distribution between the denoised images and the reference NDCT (p $>$ 0.05). In Table \ref{table4}, we show the p-values for each model across the target organs under five-fold cross-validation. A higher p-value indicates greater similarity to the radiomics features distribution of the NDCT. As shown in Table \ref{table4}, while CTformer performed well for the liver and small bowel, its performance for the aorta and right kidney was poor, exhibiting significant deviations in radiomics features distributios compared to the NDCT and even underperforming compared to LDCT. In contrast, the proposed CT-Mamba demonstrated superior performance across multiple organs, including the aorta, right kidney, and stomach. Furthermore, its radiomics features distribution closely aligned with those of the NDCT for all other target organs. Notably, for the right kidney, the CT-Mamba outperformed all other models.

\begin{table}
\scriptsize
\caption{The p-values calculated using the Wilcoxon rank-sum test for each model under five-fold cross-validation. A p-value of under 0.05 indicates statistical difference. Pairs with statistical differences are marked by *.}
\centering 
\setlength{\tabcolsep}{2pt}
\renewcommand{\arraystretch}{1.1} 
\begin{tabular}{c c c c c c}
\hline
Method  & Aorta & Right Kidney & Liver & Stomach & Small Bowel \\
\hline
LDCT    &0.3922±0.2562 &0.5373±0.3036 &0.4044±0.1834 &0.4767±0.2833 &0.3959±0.2602\\
RED-CNN  &0.4563±0.3196 &0.4743±0.3285 &0.5146±0.2938 &0.5172±0.1997 &0.7143±0.2599\\
Uformer &0.4376±0.3445 &0.5783±0.2718 &0.6046±0.1714 &0.3047±0.2488 &0.5771±0.2571\\
CTformer&0.3579±0.2883 &0.4624±0.2734 &\textbf{0.6815±0.1610} &0.5447±0.2712 & \textbf{0.7263±0.2022} \\
DDPM   &0.0002*±0.0003&0.0005*±0.0005&0.0001*±0.0001&0.0019*±0.0035&0.0009*±0.0014\\
VM-Unet &0.2805±0.3136 &0.4487±0.3539 &0.5497±0.2266 &0.6527±0.2339 &0.6527±0.2339\\
CT-Mamba&\textbf{0.4631±0.3179} &\textbf{0.6842±0.2898} &0.5722±0.2740 &\textbf{0.6440±0.2634} &0.5417±0.2992 \\
\hline
\end{tabular}
\label{table4} 
\end{table}

(2) \textbf{Pairwise Feature Similarity Ratio:} To quantitatively assess the similarity between the radiomics features of the target organs for all models and those of the NDCT, pairwise comparisons were performed for the all 167 radiomics features. The mean absolute error (MAE) between the features  of each model and the corresponding NDCT features \eqref{eq10}:
\begin{equation} \label{eq10}
R_i = \left| \frac{V_{\text{NDCT}} - V_i}{V_{\text{NDCT}}} \right|,
\end{equation}
where $i$ was the different models, and $V$ was the values of the selected radiomics features. A lower similarity ratio $R_i$ indicates higher similarity between the features of the model and those of the NDCT. For each model, the $R$ across the target organs were compared against $R_{\text{LDCT}}$, and the number of features with  $R_{\text{LDCT}}$ exceeding $R$ was recorded.

As shown in Table \ref{table5}, the proposed CT-Mamba demonstrated superior performance across all target organs. Notably, for the small bowel, where other models exhibited suboptimal performance, CT-Mamba achieved the best performance, further highlighting its robustness and effectiveness.
The radiomics analysis indicate that the proposed CT-Mamba effectively preserves the shape and texture features of multiple organs within the LDCT images, demonstrating its potential value for clinical applications.

\begin{table}
\scriptsize
\caption{The number of features with lower similarity ratios compared to LDCT (n/167, five-fold cross-validation).}  
\centering 
\setlength{\tabcolsep}{2pt}
\renewcommand{\arraystretch}{1.1} 
\begin{tabular}{c c c c c c}
\hline
Method  & Aorta & Right Kidney& Liver & Stomach & Small Bowel \\
\hline

RED-CNN  &42±26	        &48±32          &45±27           &68±30         &51±29        \\
Uformer &41±21          &45±22          &41±24           &79±20         &69±21        \\
CTformer&\textbf{50±24}	&\textbf{58±26} &45±7            &67±27         &61±12          \\
DDPM    &21±12	        &21±9           &17±8            &18±8          &20±4        \\
VM-Unet &47±23	        &48±23          &45±22           &74±13         &59±14          \\
CT-Mamba&43±23          &43±19          &\textbf{45±26}  &\textbf{82±17}&\textbf{75±17}  \\

\hline
\end{tabular}
\label{table5} 
\end{table}

\subsubsection{Ablation study}
To evaluate the effectiveness of the core components of CT-Mamba, we designed three ablation experiments and conducted quantitative validation on patient L506 from the Mayo dataset. Each ablation model was trained using data from the remaining nine patients.

(A1) Effectiveness of Deep NPS loss: Excluding Deep NPS loss during training.

(A2) Effectiveness of the Scanning Method: Replacing the Coherence ``Z" Scan with the following two methods: (A2.1) The classic row-column scanning strategy adopted by visual Mamba. (A2.2) The serpentine scanning strategy, which traverses alternating directions between adjacent rows or columns.

(M) Our complete model.

\begin{figure}[!ht]
\centerline{\includegraphics[width=\columnwidth]{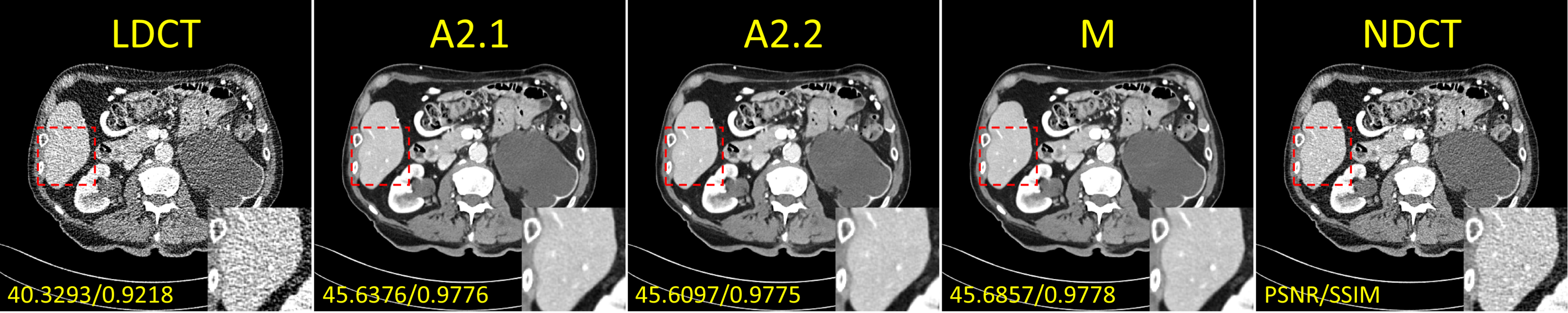}}
\caption{Comparison of prediction results under three scanning methods: (A2.1) row-column scanning, (A2.2) serpentine scanning, (M) the proposed Z-shaped scanning. The set range of the display window is a window level of 30 HU and a window width of 340 HU (i.e., a range from -140 HU to 200 HU).
}
\label{scan}
\end{figure}

\textcolor{blue}{Fig.} \ref{scan} presents the prediction results under three different scanning methods: (A2.1) row-column scanning, (A2.2) serpentine scanning, (M) the proposed Z-shaped scanning.  The results demonstrate that the Z-shaped scanning exhibits superior performance in structural modeling. In particular, it significantly enhances the clarity of vascular structures in the liver region, highlighting its advantages in maintaining spatial continuity and improving detail representation.

The quantitative results of each ablation experiment are summarized in Table. \ref{table3}. The results indicate that the complete CT-Mamba model achieves the best performance across all metrics, further validating the significant contribution of each component to the overall performance of model.

\begin{figure}[!ht]
\centerline{\includegraphics[width=\columnwidth]{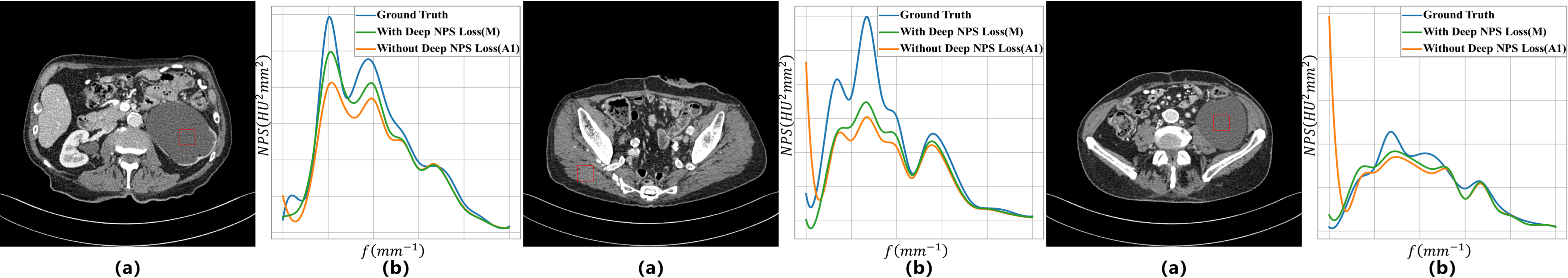}}
\caption{(a) Three representative CT slices from patient L506 in the Mayo dataset, with the red dashed box indicating the uniform ROI used for NPS analysis. (b) Radial NPS comparison of the corresponding ROIs: the green curve represents the complete model with Deep NPS loss (M), the orange curve represents the model without Deep NPS loss (A1), and the blue curve denotes the ground truth.}
\label{fig11}
\end{figure}

To further investigate the impact of the Deep NPS loss on noise texture distribution in CT-Mamba, we conducted a radial NPS analysis on a uniform ROI with a size of 32×32, as marked in \textcolor{blue}{Fig. \ref{fig11}(a)}. The corresponding radial NPS of the ROI are shown in \textcolor{blue}{Fig. \ref{fig11}(b)}. By comparing the radial NPS of the ROI under conditions without (A1, orange curve) and with (M, green curve) the Deep NPS loss, the regulatory effect of this loss on noise can be more intuitively observed. The results indicate that the Deep NPS loss effectively guides the model to generate a noise distribution more consistent with that of NDCT. This is reflected in the radial NPS, where both the frequency distribution trend and amplitude response are closer to the ground truth (blue curve), with particularly significant improvements observed in the low-frequency range. This demonstrates that the Deep NPS loss effectively guides the network to produce noise textures more consistent with NDCT, validating its effectiveness in regulating noise structure.

\begin{table}
\small
\caption{Quantitative comparison of different ablation experiments for patient (L506) in the Mayo test dataset.}
\centering 
\setlength{\tabcolsep}{5pt}
\renewcommand{\arraystretch}{0.9} 
\begin{tabular}{c c c c c}
\hline
Method   & PSNR $\uparrow$ & SSIM $\uparrow$ & RMSE $\downarrow$ & VIF $\uparrow$\\
\hline
LDCT     & 42.3509 & 0.9471 & 31.9809 & 0.2965 \\
A1       & 47.0101 & 0.9830 & 18.5470 & 0.3679 \\
A2.1       & 47.2245 & 0.9841 & 18.1040 & 0.3728 \\
A2.2       & 47.1834   & 0.9840   & 18.1874  &  0.3711   \\
\textbf{M}&\textbf{47.2825}&\textbf{0.9842}&\textbf{17.9850}&\textbf{0.3735}\\
\hline
\end{tabular}
\label{table3} 
\end{table}

\section{Conclusions}
In this paper, we propose a hybrid convolutional state-space model, CT-Mamba, for LDCT image denoising. The model combines the multi-scale analysis capability of wavelet transform, the powerful local feature extraction ability of CNN and the long-range dependent modeling advantage of Mamba, so that it can fully capture the details and global information in LDCT images. We propose a new CZSS module, which adopts a spatially coherence ``Z" Scan method, which effectively maintains the spatial continuity between adjacent pixels of the image, and further enhances the detail preservation and noise reduction capabilities of the model. Additionally, to the best of our knowledge, this is the first study to use NPS to guide deep learning in LDCT denoising tasks. We designed a deep NPS-loss driven by Mamba, aiming to ensure that the denoised image accurately restores the noise texture distribution of the NDCT image as much as possible, thereby improving the overall image quality and diagnostic value. We evaluated CT-Mamba on multiple datasets, and the experimental results show that our model performs excellently in denoising effectiveness, noise texture preservation, and radiomic feature restoration, showing potential to become a representative method of the Mamba framework for LDCT denoising tasks.

\section*{CRediT authorship contribution statement}
\textbf{Linxuan Li:} Writing - Original Draft, Methodology, Software, Conceptualization, Investigation. \textbf{Wenjia Wei:} Writing - Original Draft, Software, Methodology. \textbf{Luyao Yang:} Writing - Original Draft, Supervision, Validation. \textbf{Wenwen Zhang:} Visualization. \textbf{Jiashu Dong:} Data curation. \textbf{Yahua Liu:} Resources. 
\textbf{Hongshi Huang:} Resources. \textbf{Wei Zhao:} Writing – review and editing, Funding acquisition, Methodology, Conceptualization, Investigation, Supervision.

\section*{Acknowledgments}
This work was supported in part by the Natural Science Foundation
of Zhejiang Province (No. LZ23A050002), Natural Science Foundation
of Beijing (No. L246073), 
Natural Science Foundation of China (No. 12175012), the “111 center” (No. B20065), and the Fundamental Research Funds for the Central Universities, China.

\bibliographystyle{model2-names.bst}\biboptions{authoryear}
\flushend

\bibliography{reference}

\begin{thebibliography}{61}
\expandafter\ifx\csname natexlab\endcsname\relax\def\natexlab#1{#1}\fi
\providecommand{\url}[1]{\texttt{#1}}
\providecommand{\href}[2]{#2}
\providecommand{\path}[1]{#1}
\providecommand{\DOIprefix}{doi:}
\providecommand{\ArXivprefix}{arXiv:}
\providecommand{\URLprefix}{URL: }
\providecommand{\Pubmedprefix}{pmid:}
\providecommand{\doi}[1]{\href{http://dx.doi.org/#1}{\path{#1}}}
\providecommand{\Pubmed}[1]{\href{pmid:#1}{\path{#1}}}
\providecommand{\bibinfo}[2]{#2}
\ifx\xfnm\relax \def\xfnm[#1]{\unskip,\space#1}\fi
\bibitem[{Balda et~al.(2012)Balda, Hornegger and Heismann}]{balda2012ray}
\bibinfo{author}{Balda, M.}, \bibinfo{author}{Hornegger, J.},
  \bibinfo{author}{Heismann, B.}, \bibinfo{year}{2012}.
\newblock \bibinfo{title}{Ray contribution masks for structure adaptive
  sinogram filtering}.
\newblock \bibinfo{journal}{IEEE Trans. Med. Imaging} \bibinfo{volume}{31},
  \bibinfo{pages}{1228--1239}.
\bibitem[{Bosch~de Basea~Gomez et~al.(2023)Bosch~de Basea~Gomez, Thierry-Chef,
  Harbron, Hauptmann, Byrnes, Bernier, Le~Cornet, Dabin, Ferro, Istad
  et~al.}]{bosch2023risk}
\bibinfo{author}{Bosch~de Basea~Gomez, M.}, \bibinfo{author}{Thierry-Chef, I.},
  \bibinfo{author}{Harbron, R.}, \bibinfo{author}{Hauptmann, M.},
  \bibinfo{author}{Byrnes, G.}, \bibinfo{author}{Bernier, M.O.},
  \bibinfo{author}{Le~Cornet, L.}, \bibinfo{author}{Dabin, J.},
  \bibinfo{author}{Ferro, G.}, \bibinfo{author}{Istad, T.S.}, et~al.,
  \bibinfo{year}{2023}.
\newblock \bibinfo{title}{Risk of hematological malignancies from ct radiation
  exposure in children, adolescents and young adults}.
\newblock \bibinfo{journal}{Nat. Med.} \bibinfo{volume}{29},
  \bibinfo{pages}{3111--3119}.
\bibitem[{Chen et~al.(2017)Chen, Zhang, Kalra, Lin, Chen, Liao, Zhou and
  Wang}]{chen2017low}
\bibinfo{author}{Chen, H.}, \bibinfo{author}{Zhang, Y.},
  \bibinfo{author}{Kalra, M.K.}, \bibinfo{author}{Lin, F.},
  \bibinfo{author}{Chen, Y.}, \bibinfo{author}{Liao, P.},
  \bibinfo{author}{Zhou, J.}, \bibinfo{author}{Wang, G.}, \bibinfo{year}{2017}.
\newblock \bibinfo{title}{Low-dose ct with a residual encoder-decoder
  convolutional neural network}.
\newblock \bibinfo{journal}{IEEE Trans. Med. Imaging} \bibinfo{volume}{36},
  \bibinfo{pages}{2524--2535}.
\bibitem[{Chen et~al.(2014)Chen, Shi, Feng, Yang, Shu, Luo, Coatrieux and
  Chen}]{chen2014artifact}
\bibinfo{author}{Chen, Y.}, \bibinfo{author}{Shi, L.}, \bibinfo{author}{Feng,
  Q.}, \bibinfo{author}{Yang, J.}, \bibinfo{author}{Shu, H.},
  \bibinfo{author}{Luo, L.}, \bibinfo{author}{Coatrieux, J.L.},
  \bibinfo{author}{Chen, W.}, \bibinfo{year}{2014}.
\newblock \bibinfo{title}{Artifact suppressed dictionary learning for low-dose
  ct image processing}.
\newblock \bibinfo{journal}{IEEE Trans. Med. Imaging} \bibinfo{volume}{33},
  \bibinfo{pages}{2271--2292}.
\bibitem[{Dang et~al.(2024)Dang, Nguyen and Tiulpin}]{dang2024log}
\bibinfo{author}{Dang, T.D.Q.}, \bibinfo{author}{Nguyen, H.H.},
  \bibinfo{author}{Tiulpin, A.}, \bibinfo{year}{2024}.
\newblock \bibinfo{title}{Log-vmamba: Local-global vision mamba for medical
  image segmentation}, in: \bibinfo{booktitle}{Proceedings of the Asian
  Conference on Computer Vision}, pp. \bibinfo{pages}{548--565}.
\bibitem[{Dosovitskiy et~al.(2021)Dosovitskiy, Beyer, Kolesnikov, Weissenborn,
  Zhai, Unterthiner, Dehghani, Minderer, Heigold, Gelly, Uszkoreit and
  Houlsby}]{dosovitskiy2020image}
\bibinfo{author}{Dosovitskiy, A.}, \bibinfo{author}{Beyer, L.},
  \bibinfo{author}{Kolesnikov, A.}, \bibinfo{author}{Weissenborn, D.},
  \bibinfo{author}{Zhai, X.}, \bibinfo{author}{Unterthiner, T.},
  \bibinfo{author}{Dehghani, M.}, \bibinfo{author}{Minderer, M.},
  \bibinfo{author}{Heigold, G.}, \bibinfo{author}{Gelly, S.},
  \bibinfo{author}{Uszkoreit, J.}, \bibinfo{author}{Houlsby, N.},
  \bibinfo{year}{2021}.
\newblock \bibinfo{title}{An image is worth 16x16 words: Transformers for image
  recognition at scale}, in: \bibinfo{booktitle}{International Conference on
  Learning Representations}.
\bibitem[{Eulig et~al.(2024)Eulig, Ommer and
  Kachelrie{\ss}}]{eulig2024benchmarking}
\bibinfo{author}{Eulig, E.}, \bibinfo{author}{Ommer, B.},
  \bibinfo{author}{Kachelrie{\ss}, M.}, \bibinfo{year}{2024}.
\newblock \bibinfo{title}{Benchmarking deep learning-based low-dose ct image
  denoising algorithms}.
\newblock \bibinfo{journal}{Med. Phys.} .
\bibitem[{Fu et~al.(2024)Fu, Xiong, Lu and Zhou}]{fu2024ssumamba}
\bibinfo{author}{Fu, G.}, \bibinfo{author}{Xiong, F.}, \bibinfo{author}{Lu,
  J.}, \bibinfo{author}{Zhou, J.}, \bibinfo{year}{2024}.
\newblock \bibinfo{title}{Ssumamba: Spatial-spectral selective state space
  model for hyperspectral image denoising}.
\newblock \bibinfo{journal}{IEEE Trans. Geosci. Remote Sens.} .
\bibitem[{Fu et~al.(2025)Fu, Li, Lu, Guo, Shi, Tian and Hu}]{fu2025deep}
\bibinfo{author}{Fu, L.}, \bibinfo{author}{Li, L.}, \bibinfo{author}{Lu, B.},
  \bibinfo{author}{Guo, X.}, \bibinfo{author}{Shi, X.}, \bibinfo{author}{Tian,
  J.}, \bibinfo{author}{Hu, Z.}, \bibinfo{year}{2025}.
\newblock \bibinfo{title}{Deep equilibrium unfolding learning for noise
  estimation and removal in optical molecular imaging}.
\newblock \bibinfo{journal}{Comput. Med. Imaging Graphics}
  \bibinfo{volume}{120}, \bibinfo{pages}{102492}.
\bibitem[{Gholizadeh-Ansari et~al.(2020)Gholizadeh-Ansari, Alirezaie and
  Babyn}]{gholizadeh2020deep}
\bibinfo{author}{Gholizadeh-Ansari, M.}, \bibinfo{author}{Alirezaie, J.},
  \bibinfo{author}{Babyn, P.}, \bibinfo{year}{2020}.
\newblock \bibinfo{title}{Deep learning for low-dose ct denoising using
  perceptual loss and edge detection layer}.
\newblock \bibinfo{journal}{J DIGIT IMAGING} \bibinfo{volume}{33},
  \bibinfo{pages}{504--515}.
\bibitem[{Gu and Dao(2024)}]{gu2024mamba}
\bibinfo{author}{Gu, A.}, \bibinfo{author}{Dao, T.}, \bibinfo{year}{2024}.
\newblock \bibinfo{title}{Mamba: Linear-time sequence modeling with selective
  state spaces}.
\newblock \bibinfo{journal}{First Conference on Language Modeling} .
\bibitem[{Gu et~al.(2022)Gu, Goel and R\'e}]{gu2022efficiently}
\bibinfo{author}{Gu, A.}, \bibinfo{author}{Goel, K.}, \bibinfo{author}{R\'e,
  C.}, \bibinfo{year}{2022}.
\newblock \bibinfo{title}{Efficiently modeling long sequences with structured
  state spaces}.
\newblock \bibinfo{journal}{The International Conference on Learning
  Representations ({ICLR})} .
\bibitem[{Guo et~al.(2023)Guo, Yong, Zhang, Ma and Zhang}]{guo2023spatial}
\bibinfo{author}{Guo, S.}, \bibinfo{author}{Yong, H.}, \bibinfo{author}{Zhang,
  X.}, \bibinfo{author}{Ma, J.}, \bibinfo{author}{Zhang, L.},
  \bibinfo{year}{2023}.
\newblock \bibinfo{title}{Spatial-frequency attention for image denoising}.
\newblock \bibinfo{journal}{arXiv preprint arXiv:2302.13598} .
\bibitem[{He et~al.(2016)He, Zhang, Ren and Sun}]{he2016deep}
\bibinfo{author}{He, K.}, \bibinfo{author}{Zhang, X.}, \bibinfo{author}{Ren,
  S.}, \bibinfo{author}{Sun, J.}, \bibinfo{year}{2016}.
\newblock \bibinfo{title}{Deep residual learning for image recognition}, in:
  \bibinfo{booktitle}{Proceedings of the IEEE conference on computer vision and
  pattern recognition}, pp. \bibinfo{pages}{770--778}.
\bibitem[{Ho et~al.(2020)Ho, Jain and Abbeel}]{ho2020denoising}
\bibinfo{author}{Ho, J.}, \bibinfo{author}{Jain, A.}, \bibinfo{author}{Abbeel,
  P.}, \bibinfo{year}{2020}.
\newblock \bibinfo{title}{Denoising diffusion probabilistic models}.
\newblock \bibinfo{journal}{Adv. Neural Inf. Process. Syst.}
  \bibinfo{volume}{33}, \bibinfo{pages}{6840--6851}.
\bibitem[{Huang et~al.(2024)Huang, Zhong and Wei}]{huangnew}
\bibinfo{author}{Huang, J.}, \bibinfo{author}{Zhong, A.}, \bibinfo{author}{Wei,
  Y.}, \bibinfo{year}{2024}.
\newblock \bibinfo{title}{A new visual state space model for low-dose ct
  denoising}.
\newblock \bibinfo{journal}{Med. Phys.} .
\bibitem[{Iqbal et~al.(2020)Iqbal, Shahzad, Rafiq, Mustafa and
  Ma}]{iqbal2020deep}
\bibinfo{author}{Iqbal, I.}, \bibinfo{author}{Shahzad, G.},
  \bibinfo{author}{Rafiq, N.}, \bibinfo{author}{Mustafa, G.},
  \bibinfo{author}{Ma, J.}, \bibinfo{year}{2020}.
\newblock \bibinfo{title}{Deep learning-based automated detection of human knee
  joint's synovial fluid from magnetic resonance images with transfer
  learning}.
\newblock \bibinfo{journal}{IET Image Process.} \bibinfo{volume}{14},
  \bibinfo{pages}{1990--1998}.
\bibitem[{Iqbal et~al.(2021)Iqbal, Younus, Walayat, Kakar and
  Ma}]{iqbal2021automated}
\bibinfo{author}{Iqbal, I.}, \bibinfo{author}{Younus, M.},
  \bibinfo{author}{Walayat, K.}, \bibinfo{author}{Kakar, M.U.},
  \bibinfo{author}{Ma, J.}, \bibinfo{year}{2021}.
\newblock \bibinfo{title}{Automated multi-class classification of skin lesions
  through deep convolutional neural network with dermoscopic images}.
\newblock \bibinfo{journal}{Comput. Med. Imaging Graph.} \bibinfo{volume}{88},
  \bibinfo{pages}{101843}.
\bibitem[{Isensee et~al.(2021)Isensee, Jaeger, Kohl, Petersen and
  Maier-Hein}]{isensee2021nnu}
\bibinfo{author}{Isensee, F.}, \bibinfo{author}{Jaeger, P.F.},
  \bibinfo{author}{Kohl, S.A.}, \bibinfo{author}{Petersen, J.},
  \bibinfo{author}{Maier-Hein, K.H.}, \bibinfo{year}{2021}.
\newblock \bibinfo{title}{nnu-net: a self-configuring method for deep
  learning-based biomedical image segmentation}.
\newblock \bibinfo{journal}{Nat. Methods} \bibinfo{volume}{18},
  \bibinfo{pages}{203--211}.
\bibitem[{Ji et~al.(2022)Ji, Bai, Ge, Yang, Zhu, Zhang, Li, Zhanng, Ma, Wan
  et~al.}]{ji2022amos}
\bibinfo{author}{Ji, Y.}, \bibinfo{author}{Bai, H.}, \bibinfo{author}{Ge, C.},
  \bibinfo{author}{Yang, J.}, \bibinfo{author}{Zhu, Y.},
  \bibinfo{author}{Zhang, R.}, \bibinfo{author}{Li, Z.},
  \bibinfo{author}{Zhanng, L.}, \bibinfo{author}{Ma, W.}, \bibinfo{author}{Wan,
  X.}, et~al., \bibinfo{year}{2022}.
\newblock \bibinfo{title}{Amos: A large-scale abdominal multi-organ benchmark
  for versatile medical image segmentation}.
\newblock \bibinfo{journal}{Advances in neural information processing systems}
  \bibinfo{volume}{35}, \bibinfo{pages}{36722--36732}.
\bibitem[{Kang et~al.(2013)Kang, Slomka, Nakazato, Woo, Berman, Kuo and
  Dey}]{kang2013image}
\bibinfo{author}{Kang, D.}, \bibinfo{author}{Slomka, P.},
  \bibinfo{author}{Nakazato, R.}, \bibinfo{author}{Woo, J.},
  \bibinfo{author}{Berman, D.S.}, \bibinfo{author}{Kuo, C.C.J.},
  \bibinfo{author}{Dey, D.}, \bibinfo{year}{2013}.
\newblock \bibinfo{title}{Image denoising of low-radiation dose coronary ct
  angiography by an adaptive block-matching 3d algorithm}, in:
  \bibinfo{booktitle}{Medical Imaging 2013: Image Processing},
  \bibinfo{organization}{SPIE}. pp. \bibinfo{pages}{671--676}.
\bibitem[{Kebaili et~al.(2025)Kebaili, Lapuyade-Lahorgue, Vera and
  Ruan}]{kebaili2025multi}
\bibinfo{author}{Kebaili, A.}, \bibinfo{author}{Lapuyade-Lahorgue, J.},
  \bibinfo{author}{Vera, P.}, \bibinfo{author}{Ruan, S.}, \bibinfo{year}{2025}.
\newblock \bibinfo{title}{Multi-modal mri synthesis with conditional latent
  diffusion models for data augmentation in tumor segmentation}.
\newblock \bibinfo{journal}{Comput. Med. Imaging Graph.} \bibinfo{volume}{123},
  \bibinfo{pages}{102532}.
\bibitem[{Ko et~al.(2024)Ko, Song, Baek and Shim}]{ko2024adapting}
\bibinfo{author}{Ko, Y.}, \bibinfo{author}{Song, S.}, \bibinfo{author}{Baek,
  J.}, \bibinfo{author}{Shim, H.}, \bibinfo{year}{2024}.
\newblock \bibinfo{title}{Adapting low-dose ct denoisers for texture
  preservation using zero-shot local noise-level matching}.
\newblock \bibinfo{journal}{Med. Phys.} \bibinfo{volume}{51},
  \bibinfo{pages}{4181--4200}.
\bibitem[{Kumar and Kurmi(2022)}]{kumar2022cnn}
\bibinfo{author}{Kumar, S.}, \bibinfo{author}{Kurmi, Y.}, \bibinfo{year}{2022}.
\newblock \bibinfo{title}{Cnn-based denoising system for the image quality
  enhancement}.
\newblock \bibinfo{journal}{Multimed. Tools Appl.} \bibinfo{volume}{81},
  \bibinfo{pages}{20147--20174}.
\bibitem[{Kurmi et~al.(2024)Kurmi, Viswanathan and Zu}]{kurmi2024enhancing}
\bibinfo{author}{Kurmi, Y.}, \bibinfo{author}{Viswanathan, M.},
  \bibinfo{author}{Zu, Z.}, \bibinfo{year}{2024}.
\newblock \bibinfo{title}{Enhancing snr in cest imaging: A deep learning
  approach with a denoising convolutional autoencoder}.
\newblock \bibinfo{journal}{Magn. Reson. Med.} \bibinfo{volume}{92},
  \bibinfo{pages}{2404--2419}.
\bibitem[{Kyung et~al.(2024)Kyung, Won, Pak, Kim, Lee, Park, Hong and
  Kim}]{kyung2024generative}
\bibinfo{author}{Kyung, S.}, \bibinfo{author}{Won, J.}, \bibinfo{author}{Pak,
  S.}, \bibinfo{author}{Kim, S.}, \bibinfo{author}{Lee, S.},
  \bibinfo{author}{Park, K.}, \bibinfo{author}{Hong, G.S.},
  \bibinfo{author}{Kim, N.}, \bibinfo{year}{2024}.
\newblock \bibinfo{title}{Generative adversarial network with robust
  discriminator through multi-task learning for low-dose ct denoising}.
\newblock \bibinfo{journal}{IEEE Trans. Med. Imaging} .
\bibitem[{LeCun et~al.(1995)LeCun, Bengio et~al.}]{lecun1995convolutional}
\bibinfo{author}{LeCun, Y.}, \bibinfo{author}{Bengio, Y.}, et~al.,
  \bibinfo{year}{1995}.
\newblock \bibinfo{title}{Convolutional networks for images, speech, and time
  series}.
\newblock \bibinfo{journal}{The handbook of brain theory and neural networks}
  \bibinfo{volume}{3361}, \bibinfo{pages}{1995}.
\bibitem[{Li et~al.(2025a)Li, Wei, Lu, Zhang, Zhang and Zhao}]{10990033}
\bibinfo{author}{Li, L.}, \bibinfo{author}{Wei, W.}, \bibinfo{author}{Lu, Y.},
  \bibinfo{author}{Zhang, W.}, \bibinfo{author}{Zhang, Y.},
  \bibinfo{author}{Zhao, W.}, \bibinfo{year}{2025}a.
\newblock \bibinfo{title}{Bsonet: Deep learning solution for optimizing image
  quality of portable backscatter imaging systems}.
\newblock \bibinfo{journal}{IEEE Trans. Comput. Imaging} \bibinfo{volume}{11},
  \bibinfo{pages}{650--662}.
\bibitem[{Li et~al.(2025b)Li, Zhang, Li, Wang and Zhao}]{li2025ddoct}
\bibinfo{author}{Li, L.}, \bibinfo{author}{Zhang, Z.}, \bibinfo{author}{Li,
  Y.}, \bibinfo{author}{Wang, Y.}, \bibinfo{author}{Zhao, W.},
  \bibinfo{year}{2025}b.
\newblock \bibinfo{title}{Ddoct: Morphology preserved dual-domain joint
  optimization for fast sparse-view low-dose ct imaging}.
\newblock \bibinfo{journal}{Med. Image Anal.} \bibinfo{volume}{101},
  \bibinfo{pages}{103420}.
\bibitem[{Li et~al.(2022)Li, Li, Li, Wu, Dong, Zhao, Qiang and
  Aftab}]{li2022low}
\bibinfo{author}{Li, Q.}, \bibinfo{author}{Li, S.}, \bibinfo{author}{Li, R.},
  \bibinfo{author}{Wu, W.}, \bibinfo{author}{Dong, Y.}, \bibinfo{author}{Zhao,
  J.}, \bibinfo{author}{Qiang, Y.}, \bibinfo{author}{Aftab, R.},
  \bibinfo{year}{2022}.
\newblock \bibinfo{title}{Low-dose computed tomography image reconstruction via
  a multistage convolutional neural network with autoencoder perceptual loss
  network}.
\newblock \bibinfo{journal}{QUANT IMAG MED SURG} \bibinfo{volume}{12},
  \bibinfo{pages}{1929}.
\bibitem[{Li et~al.(2014)Li, Yu, Trzasko, Lake, Blezek, Fletcher, McCollough
  and Manduca}]{li2014adaptive}
\bibinfo{author}{Li, Z.}, \bibinfo{author}{Yu, L.}, \bibinfo{author}{Trzasko,
  J.D.}, \bibinfo{author}{Lake, D.S.}, \bibinfo{author}{Blezek, D.J.},
  \bibinfo{author}{Fletcher, J.G.}, \bibinfo{author}{McCollough, C.H.},
  \bibinfo{author}{Manduca, A.}, \bibinfo{year}{2014}.
\newblock \bibinfo{title}{Adaptive nonlocal means filtering based on local
  noise level for ct denoising}.
\newblock \bibinfo{journal}{Med. Phys.} \bibinfo{volume}{41},
  \bibinfo{pages}{011908}.
\bibitem[{Lv et~al.(2024)Lv, Xie, Zhang, Liu, Zhang, Qu, Zhao and
  Xu}]{lv2024qualitative}
\bibinfo{author}{Lv, T.}, \bibinfo{author}{Xie, C.}, \bibinfo{author}{Zhang,
  Y.}, \bibinfo{author}{Liu, Y.}, \bibinfo{author}{Zhang, G.},
  \bibinfo{author}{Qu, B.}, \bibinfo{author}{Zhao, W.}, \bibinfo{author}{Xu,
  S.}, \bibinfo{year}{2024}.
\newblock \bibinfo{title}{A qualitative study of improving megavoltage computed
  tomography image quality and maintaining dose accuracy using cyclegan-based
  image synthesis}.
\newblock \bibinfo{journal}{Med. Phys.} \bibinfo{volume}{51},
  \bibinfo{pages}{394--406}.
\bibitem[{Ma et~al.(2025)Ma, Zou, Fang, Luo, Wang, Dong, Li, Wang, Dong, Tian
  et~al.}]{ma2025convergent}
\bibinfo{author}{Ma, X.}, \bibinfo{author}{Zou, M.}, \bibinfo{author}{Fang,
  X.}, \bibinfo{author}{Luo, G.}, \bibinfo{author}{Wang, W.},
  \bibinfo{author}{Dong, S.}, \bibinfo{author}{Li, X.}, \bibinfo{author}{Wang,
  K.}, \bibinfo{author}{Dong, Q.}, \bibinfo{author}{Tian, Y.}, et~al.,
  \bibinfo{year}{2025}.
\newblock \bibinfo{title}{Convergent--diffusion denoising model for
  multi-scenario ct image reconstruction}.
\newblock \bibinfo{journal}{Comput. Med. Imaging Graph.} \bibinfo{volume}{120},
  \bibinfo{pages}{102491}.
\bibitem[{Manduca et~al.(2009)Manduca, Yu, Trzasko, Khaylova, Kofler,
  McCollough and Fletcher}]{manduca2009projection}
\bibinfo{author}{Manduca, A.}, \bibinfo{author}{Yu, L.},
  \bibinfo{author}{Trzasko, J.D.}, \bibinfo{author}{Khaylova, N.},
  \bibinfo{author}{Kofler, J.M.}, \bibinfo{author}{McCollough, C.M.},
  \bibinfo{author}{Fletcher, J.G.}, \bibinfo{year}{2009}.
\newblock \bibinfo{title}{Projection space denoising with bilateral filtering
  and ct noise modeling for dose reduction in ct}.
\newblock \bibinfo{journal}{Med. Phys.} \bibinfo{volume}{36},
  \bibinfo{pages}{4911--4919}.
\bibitem[{McCollough et~al.(2017)McCollough, Bartley, Carter, Chen, Drees,
  Edwards, Holmes~III, Huang, Khan, Leng et~al.}]{mccollough2017low}
\bibinfo{author}{McCollough, C.H.}, \bibinfo{author}{Bartley, A.C.},
  \bibinfo{author}{Carter, R.E.}, \bibinfo{author}{Chen, B.},
  \bibinfo{author}{Drees, T.A.}, \bibinfo{author}{Edwards, P.},
  \bibinfo{author}{Holmes~III, D.R.}, \bibinfo{author}{Huang, A.E.},
  \bibinfo{author}{Khan, F.}, \bibinfo{author}{Leng, S.}, et~al.,
  \bibinfo{year}{2017}.
\newblock \bibinfo{title}{Low-dose ct for the detection and classification of
  metastatic liver lesions: results of the 2016 low dose ct grand challenge}.
\newblock \bibinfo{journal}{Med. Phys.} \bibinfo{volume}{44},
  \bibinfo{pages}{e339--e352}.
\bibitem[{Meng et~al.(2024)Meng, Wang, Zhu, Tao, Mao, Liao, Bian, Zeng and
  Ma}]{meng2024ddt}
\bibinfo{author}{Meng, M.}, \bibinfo{author}{Wang, Y.}, \bibinfo{author}{Zhu,
  M.}, \bibinfo{author}{Tao, X.}, \bibinfo{author}{Mao, Z.},
  \bibinfo{author}{Liao, J.}, \bibinfo{author}{Bian, Z.},
  \bibinfo{author}{Zeng, D.}, \bibinfo{author}{Ma, J.}, \bibinfo{year}{2024}.
\newblock \bibinfo{title}{Ddt-net: Dose-agnostic dual-task transfer network for
  simultaneous low-dose ct denoising and simulation}.
\newblock \bibinfo{journal}{IEEE J. Biomed. Health. Inf.} .
\bibitem[{Ming et~al.(2020)Ming, Yi, Zhang and Li}]{ming2020low}
\bibinfo{author}{Ming, J.}, \bibinfo{author}{Yi, B.}, \bibinfo{author}{Zhang,
  Y.}, \bibinfo{author}{Li, H.}, \bibinfo{year}{2020}.
\newblock \bibinfo{title}{Low-dose ct image denoising using classification
  densely connected residual network}.
\newblock \bibinfo{journal}{KSII Trans. Internet Inf. Syst.}
  \bibinfo{volume}{14}, \bibinfo{pages}{2480--2496}.
\bibitem[{Patwari et~al.(2023)Patwari, Gutjahr, Marcus, Thali, Calvarons,
  Raupach and Maier}]{patwari2023reducing}
\bibinfo{author}{Patwari, M.}, \bibinfo{author}{Gutjahr, R.},
  \bibinfo{author}{Marcus, R.}, \bibinfo{author}{Thali, Y.},
  \bibinfo{author}{Calvarons, A.F.}, \bibinfo{author}{Raupach, R.},
  \bibinfo{author}{Maier, A.}, \bibinfo{year}{2023}.
\newblock \bibinfo{title}{Reducing the risk of hallucinations with
  interpretable deep learning models for low-dose ct denoising: comparative
  performance analysis}.
\newblock \bibinfo{journal}{Phys. Med. Biol.} \bibinfo{volume}{68},
  \bibinfo{pages}{19LT01}.
\bibitem[{Redmon(2016)}]{redmon2016you}
\bibinfo{author}{Redmon, J.}, \bibinfo{year}{2016}.
\newblock \bibinfo{title}{You only look once: Unified, real-time object
  detection}, in: \bibinfo{booktitle}{Proceedings of the IEEE conference on
  computer vision and pattern recognition}.
\bibitem[{Rombach et~al.(2022)Rombach, Blattmann, Lorenz, Esser and
  Ommer}]{Rombach_2022_CVPR}
\bibinfo{author}{Rombach, R.}, \bibinfo{author}{Blattmann, A.},
  \bibinfo{author}{Lorenz, D.}, \bibinfo{author}{Esser, P.},
  \bibinfo{author}{Ommer, B.}, \bibinfo{year}{2022}.
\newblock \bibinfo{title}{High-resolution image synthesis with latent diffusion
  models}, in: \bibinfo{booktitle}{Proceedings of the IEEE/CVF Conference on
  Computer Vision and Pattern Recognition (CVPR)}, pp.
  \bibinfo{pages}{10684--10695}.
\bibitem[{Ronneberger et~al.(2015)Ronneberger, Fischer and
  Brox}]{ronneberger2015u}
\bibinfo{author}{Ronneberger, O.}, \bibinfo{author}{Fischer, P.},
  \bibinfo{author}{Brox, T.}, \bibinfo{year}{2015}.
\newblock \bibinfo{title}{U-net: Convolutional networks for biomedical image
  segmentation}, in: \bibinfo{booktitle}{Medical image computing and
  computer-assisted intervention--MICCAI 2015: 18th international conference,
  Munich, Germany, October 5-9, 2015, proceedings, part III 18},
  \bibinfo{organization}{Springer}. pp. \bibinfo{pages}{234--241}.
\bibitem[{Ruan and Xiang(2024)}]{ruan2024vm}
\bibinfo{author}{Ruan, J.}, \bibinfo{author}{Xiang, S.}, \bibinfo{year}{2024}.
\newblock \bibinfo{title}{Vm-unet: Vision mamba unet for medical image
  segmentation}.
\newblock \bibinfo{journal}{arXiv preprint arXiv:2402.02491} .
\bibitem[{Song et~al.(2021)Song, Meng and Ermon}]{song2021denoising}
\bibinfo{author}{Song, J.}, \bibinfo{author}{Meng, C.}, \bibinfo{author}{Ermon,
  S.}, \bibinfo{year}{2021}.
\newblock \bibinfo{title}{Denoising diffusion implicit models}, in:
  \bibinfo{booktitle}{Int. Conf. Learn. Represent.}
\bibitem[{Vaswani(2017)}]{vaswani2017attention}
\bibinfo{author}{Vaswani, A.}, \bibinfo{year}{2017}.
\newblock \bibinfo{title}{Attention is all you need}.
\newblock \bibinfo{journal}{Advances in Neural Information Processing Systems}
  .
\bibitem[{Wang et~al.(2023)Wang, Fan, Wu, Liu, Wang and Yu}]{wang2023ctformer}
\bibinfo{author}{Wang, D.}, \bibinfo{author}{Fan, F.}, \bibinfo{author}{Wu,
  Z.}, \bibinfo{author}{Liu, R.}, \bibinfo{author}{Wang, F.},
  \bibinfo{author}{Yu, H.}, \bibinfo{year}{2023}.
\newblock \bibinfo{title}{Ctformer: convolution-free token2token dilated vision
  transformer for low-dose ct denoising}.
\newblock \bibinfo{journal}{Phys. Med. Biol.} \bibinfo{volume}{68},
  \bibinfo{pages}{065012}.
\bibitem[{Wang et~al.(2006)Wang, Li, Lu and Liang}]{wang2006penalized}
\bibinfo{author}{Wang, J.}, \bibinfo{author}{Li, T.}, \bibinfo{author}{Lu, H.},
  \bibinfo{author}{Liang, Z.}, \bibinfo{year}{2006}.
\newblock \bibinfo{title}{Penalized weighted least-squares approach to sinogram
  noise reduction and image reconstruction for low-dose x-ray computed
  tomography}.
\newblock \bibinfo{journal}{IEEE Trans. Med. Imaging} \bibinfo{volume}{25},
  \bibinfo{pages}{1272--1283}.
\bibitem[{Wang and Wang(2022)}]{wang2022simulating}
\bibinfo{author}{Wang, S.}, \bibinfo{author}{Wang, A.S.}, \bibinfo{year}{2022}.
\newblock \bibinfo{title}{Simulating arbitrary dose levels and independent
  noise image pairs from a single ct scan}, in: \bibinfo{booktitle}{7th
  International Conference on Image Formation in X-Ray Computed Tomography},
  \bibinfo{organization}{SPIE}. pp. \bibinfo{pages}{460--466}.
\bibitem[{Wang et~al.(2024)Wang, Xia, Lu and Zhang}]{wang2024review}
\bibinfo{author}{Wang, T.}, \bibinfo{author}{Xia, W.}, \bibinfo{author}{Lu,
  J.}, \bibinfo{author}{Zhang, Y.}, \bibinfo{year}{2024}.
\newblock \bibinfo{title}{A review of deep learning ct reconstruction from
  incomplete projection data}.
\newblock \bibinfo{journal}{IEEE Trans. Radiat. Plasma Med. Sci.}
  \bibinfo{volume}{8}, \bibinfo{pages}{138--152}.
\bibitem[{Wang et~al.(2022)Wang, Cun, Bao, Zhou, Liu and Li}]{wang2022uformer}
\bibinfo{author}{Wang, Z.}, \bibinfo{author}{Cun, X.}, \bibinfo{author}{Bao,
  J.}, \bibinfo{author}{Zhou, W.}, \bibinfo{author}{Liu, J.},
  \bibinfo{author}{Li, H.}, \bibinfo{year}{2022}.
\newblock \bibinfo{title}{Uformer: A general u-shaped transformer for image
  restoration}, in: \bibinfo{booktitle}{Proceedings of the IEEE/CVF conference
  on computer vision and pattern recognition}, pp.
  \bibinfo{pages}{17683--17693}.
\bibitem[{Wasserthal et~al.(2023)Wasserthal, Breit, Meyer, Pradella, Hinck,
  Sauter, Heye, Boll, Cyriac, Yang et~al.}]{wasserthal2023totalsegmentator}
\bibinfo{author}{Wasserthal, J.}, \bibinfo{author}{Breit, H.C.},
  \bibinfo{author}{Meyer, M.T.}, \bibinfo{author}{Pradella, M.},
  \bibinfo{author}{Hinck, D.}, \bibinfo{author}{Sauter, A.W.},
  \bibinfo{author}{Heye, T.}, \bibinfo{author}{Boll, D.T.},
  \bibinfo{author}{Cyriac, J.}, \bibinfo{author}{Yang, S.}, et~al.,
  \bibinfo{year}{2023}.
\newblock \bibinfo{title}{Totalsegmentator: robust segmentation of 104 anatomic
  structures in ct images}.
\newblock \bibinfo{journal}{Radiology-Artificial Intelligence}
  \bibinfo{volume}{5}.
\bibitem[{Wilson et~al.(2013)Wilson, Christianson, Richard and
  Samei}]{wilson2013methodology}
\bibinfo{author}{Wilson, J.M.}, \bibinfo{author}{Christianson, O.I.},
  \bibinfo{author}{Richard, S.}, \bibinfo{author}{Samei, E.},
  \bibinfo{year}{2013}.
\newblock \bibinfo{title}{A methodology for image quality evaluation of
  advanced ct systems}.
\newblock \bibinfo{journal}{Med. Phys.} \bibinfo{volume}{40},
  \bibinfo{pages}{031908}.
\bibitem[{Xia et~al.(2025)Xia, Yan, Yang, Zhang and Tao}]{xia2025cmc}
\bibinfo{author}{Xia, J.}, \bibinfo{author}{Yan, M.}, \bibinfo{author}{Yang,
  X.}, \bibinfo{author}{Zhang, X.}, \bibinfo{author}{Tao, Z.},
  \bibinfo{year}{2025}.
\newblock \bibinfo{title}{Cmc-diffusion: Curve matching correction diffusion
  model for ldct denoising}.
\newblock \bibinfo{journal}{Biomed. Signal Process. Control}
  \bibinfo{volume}{103}, \bibinfo{pages}{107333}.
\bibitem[{Xia et~al.(2023)Xia, Zhou, Meng, Zha, Yu, Wang, Song, Wang, Tang, Xu
  et~al.}]{xia2023predicting}
\bibinfo{author}{Xia, T.y.}, \bibinfo{author}{Zhou, Z.h.},
  \bibinfo{author}{Meng, X.p.}, \bibinfo{author}{Zha, J.h.},
  \bibinfo{author}{Yu, Q.}, \bibinfo{author}{Wang, W.l.},
  \bibinfo{author}{Song, Y.}, \bibinfo{author}{Wang, Y.c.},
  \bibinfo{author}{Tang, T.y.}, \bibinfo{author}{Xu, J.}, et~al.,
  \bibinfo{year}{2023}.
\newblock \bibinfo{title}{Predicting microvascular invasion in hepatocellular
  carcinoma using ct-based radiomics model}.
\newblock \bibinfo{journal}{RADIOLOGY} \bibinfo{volume}{307},
  \bibinfo{pages}{e222729}.
\bibitem[{Xie et~al.(2024)Xie, Zhang, Chen and Tan}]{xie2024precision}
\bibinfo{author}{Xie, H.}, \bibinfo{author}{Zhang, H.}, \bibinfo{author}{Chen,
  Z.}, \bibinfo{author}{Tan, T.}, \bibinfo{year}{2024}.
\newblock \bibinfo{title}{Precision dose prediction for breast cancer patients
  undergoing imrt: The swin-umamba-channel model}.
\newblock \bibinfo{journal}{Comput. Med. Imaging Graphics}
  \bibinfo{volume}{116}, \bibinfo{pages}{102409}.
\bibitem[{Yang et~al.(2023)Yang, Zhang, Song, Hong, Xu, Zhao, Zhang, Cui and
  Yang}]{yang2023diffusion}
\bibinfo{author}{Yang, L.}, \bibinfo{author}{Zhang, Z.}, \bibinfo{author}{Song,
  Y.}, \bibinfo{author}{Hong, S.}, \bibinfo{author}{Xu, R.},
  \bibinfo{author}{Zhao, Y.}, \bibinfo{author}{Zhang, W.},
  \bibinfo{author}{Cui, B.}, \bibinfo{author}{Yang, M.H.},
  \bibinfo{year}{2023}.
\newblock \bibinfo{title}{Diffusion models: A comprehensive survey of methods
  and applications}.
\newblock \bibinfo{journal}{ACM Comput. Surv.} \bibinfo{volume}{56},
  \bibinfo{pages}{1--39}.
\bibitem[{Yang et~al.(2018)Yang, Yan, Zhang, Yu, Shi, Mou, Kalra, Zhang, Sun
  and Wang}]{yang2018low}
\bibinfo{author}{Yang, Q.}, \bibinfo{author}{Yan, P.}, \bibinfo{author}{Zhang,
  Y.}, \bibinfo{author}{Yu, H.}, \bibinfo{author}{Shi, Y.},
  \bibinfo{author}{Mou, X.}, \bibinfo{author}{Kalra, M.K.},
  \bibinfo{author}{Zhang, Y.}, \bibinfo{author}{Sun, L.},
  \bibinfo{author}{Wang, G.}, \bibinfo{year}{2018}.
\newblock \bibinfo{title}{Low-dose ct image denoising using a generative
  adversarial network with wasserstein distance and perceptual loss}.
\newblock \bibinfo{journal}{IEEE Trans. Med. Imaging} \bibinfo{volume}{37},
  \bibinfo{pages}{1348--1357}.
\bibitem[{Zhang et~al.(2024a)Zhang, Gong, Ye, Wang, Shangguan and
  Cheng}]{zhang2024review}
\bibinfo{author}{Zhang, J.}, \bibinfo{author}{Gong, W.}, \bibinfo{author}{Ye,
  L.}, \bibinfo{author}{Wang, F.}, \bibinfo{author}{Shangguan, Z.},
  \bibinfo{author}{Cheng, Y.}, \bibinfo{year}{2024}a.
\newblock \bibinfo{title}{A review of deep learning methods for denoising of
  medical low-dose ct images}.
\newblock \bibinfo{journal}{Comput. Biol. Med.} , \bibinfo{pages}{108112}.
\bibitem[{Zhang et~al.(2025)Zhang, Lyu, Li, Chen, Sun and Zhao}]{zhang2025data}
\bibinfo{author}{Zhang, W.}, \bibinfo{author}{Lyu, T.}, \bibinfo{author}{Li,
  Y.}, \bibinfo{author}{Chen, Y.}, \bibinfo{author}{Sun, B.},
  \bibinfo{author}{Zhao, W.}, \bibinfo{year}{2025}.
\newblock \bibinfo{title}{Data-driven contrast-enhanced dual-energy ct imaging
  via physically constrained attention}.
\newblock \bibinfo{journal}{IEEE Trans. Radiat. Plasma Med. Sci.} .
\bibitem[{Zhang et~al.(2024b)Zhang, Wan, Wang, Meng, Ma, Guo, Liu, Li and
  Liu}]{zhang2024multi}
\bibinfo{author}{Zhang, Y.}, \bibinfo{author}{Wan, Z.}, \bibinfo{author}{Wang,
  D.}, \bibinfo{author}{Meng, J.}, \bibinfo{author}{Ma, F.},
  \bibinfo{author}{Guo, Y.}, \bibinfo{author}{Liu, J.}, \bibinfo{author}{Li,
  G.}, \bibinfo{author}{Liu, Y.}, \bibinfo{year}{2024}b.
\newblock \bibinfo{title}{Multi-scale feature aggregation and fusion network
  with self-supervised multi-level perceptual loss for textures preserving
  low-dose ct denoising}.
\newblock \bibinfo{journal}{Phys. Med. Biol.} \bibinfo{volume}{69},
  \bibinfo{pages}{105003}.
\bibitem[{Zhang et~al.(2021)Zhang, Liang, Zhao and
  Xing}]{zhang2021noise2context}
\bibinfo{author}{Zhang, Z.}, \bibinfo{author}{Liang, X.},
  \bibinfo{author}{Zhao, W.}, \bibinfo{author}{Xing, L.}, \bibinfo{year}{2021}.
\newblock \bibinfo{title}{Noise2context: context-assisted learning 3d
  thin-layer for low-dose ct}.
\newblock \bibinfo{journal}{Med. Phys.} \bibinfo{volume}{48},
  \bibinfo{pages}{5794--5803}.
\bibitem[{Zhao et~al.(2024)Zhao, Chen, Zhang, Xiao, Bai and
  Ouyang}]{zhao2024rs}
\bibinfo{author}{Zhao, S.}, \bibinfo{author}{Chen, H.}, \bibinfo{author}{Zhang,
  X.}, \bibinfo{author}{Xiao, P.}, \bibinfo{author}{Bai, L.},
  \bibinfo{author}{Ouyang, W.}, \bibinfo{year}{2024}.
\newblock \bibinfo{title}{Rs-mamba for large remote sensing image dense
  prediction}.
\newblock \bibinfo{journal}{IEEE Trans. Geosci. Remote Sens.} .

\end{thebibliography}

\end{document}